\documentclass[a4paper,fleqn,usenatbib]{mnras}
\usepackage{amsmath}
\usepackage{epsfig} 
\usepackage{graphicx}	
\usepackage{amssymb}	

\newcommand{\be}{\begin{equation}}
\newcommand{\e}{\end{equation}}
\newcommand{\bear}{\begin{eqnarray}}
\newcommand{\ear}{\end{eqnarray}}

\newcommand{\hmpc}{{\, h^{-1}\, {\rm Mpc}}}

\def\aj{AJ}
\def\apj{ApJ}
\def\apjs{ApJS}
\def\jcap{JCAP}
\def\mnras{MNRAS}
\def\aap{A\&A}

\def\nat{Nature}      
\def\apjs{ApJS}
\def\apjl{ApJ Letters}

\title[Local Dimension in SDSS DR14] {Unravelling the Cosmic Web: An analysis of the SDSS DR14 with the Local Dimension} 

\author[Sarkar, S. and Pandey, B.] {Suman
  Sarkar$^{1}$\thanks{suman2reach@gmail.com} and Biswajit
  Pandey$^1$\thanks{biswap@visva-bharati.ac.in} \\$^1$ Department of
  Physics, Visva-Bharati University, Santiniketan, Birbhum, 731235,
  India }

 \date{\today}

 \pubyear{2018}
  
\begin{document}
\label{firstpage}
\pagerange{\pageref{firstpage}--\pageref{lastpage}}      
\maketitle
       
\begin{abstract}
  We analyze a volume limited galaxy sample from the SDSS to study the
  environments of galaxies on different length scales in the local
  Universe. We measure the local dimension of the SDSS galaxies on
  different length scales and find that the sheets or sheetlike
  structures are the most prevalent pattern in the cosmic web
  throughout the entire length scales. The abundance of sheets peaks
  at $30 \hmpc$ and they can extend upto a length scales of $90
  \hmpc$. Analyzing mock catalogues, we find that the sheets are
  non-existent beyond $30 \hmpc$ in the Poisson distributions. We find
  that the straight filaments in the SDSS galaxy distribution can
  extend only upto a length scale of $30 \hmpc$. Our results indicate
  that the environment of a galaxy exhibits a gradual transition
  towards higher local dimension with increasing length scales finally
  approaching a nearly homogeneous network on large scales. We compare
  our findings with a semi analytic galaxy catalogue from the
  Millennium Run simulation which are in fairly good agreement with
  the observations. We also test the effects of the number density of
  the sample and the cut-off in the goodness of fit which shows that
  the results are nearly independent of these factors. Finally we
  apply the method to a set of simulations of the segment Cox process
  and find that it can characterize such distributions.

\end{abstract}
       

\section{Introduction}
Understanding the formation and evolution of the cosmic web remains
one of the most fascinating and challenging problems in cosmology. The
first observational hint of the existence of the cosmic web came
through several early redshift surveys
\citep{chincarini,gregory,einasto} which was later confirmed
\citep{delapparent} by the surveys like CfA \citep{davis} and LCRS
\citep{shectman}. The modern redshift surveys like the 2dFGRS
\citep{colless} and the SDSS \citep{york} have now revealed the cosmic
web in its full glory. The cosmic web is a network of galaxies
spanning the entire Universe. The network comprises of several
distinct morphological components such as clusters, filaments and
sheets which are interconnected in a complex manner and are
encompassed by voids of numerous sizes. The galaxies form and evolve
in different environments inside the cosmic web and the different
morphological components provide unique environments for galaxy
formation and evolution.

The first theoretical insight into the formation of the cosmic web was
provided by the seminal work of \citet{zeldovich} which showed how the
successive collapse of an overdense region along its longest, medium
and shortest axis would produce spatial patterns like sheets,
filaments and clusters respectively. Characterizing these spatial
patterns in the cosmic web is an important step towards understanding
the galaxy formation and evolution in the Universe. A large number of
statistical tools have been designed for this purpose. The
  percolation analysis \citep{shandarin1983, einasto1}, the genus
  \citep{Gott1, appleby}, the Minkowski functionals \citep{mecke1994,
    wiegand, fang}, the Shapefinders \citep{Sahni1998, bharadwaj2004,
    pandey1, bag}, the minimal spanning tree \citep{barrow,lares}, the
  statistics of maxima and saddle points \citep{colombi, ansari}, the
  multiscale morphology filter based on the Hessian of the density
  field \citep{arag, arag1}, the skeleton formalism \citep{novikov,
    sous}, the local dimension \citep{sarkar, sarkar1} and the Origami
  approximation \citep{neyrinck, neyrinck1} are to name a few. Each
of these different statistical measures captures some aspects of the
cosmic web. But a comprehensive measure of the cosmic web is still
awaited. Presently, developing effective tools for the quantification
of the cosmic web is an active area of research.

The different structural elements of the cosmic web are characterized
by their density and geometry. The galaxy clusters located at the
nodes where the filaments intersect, are known to be the densest
regions in the cosmic web followed by the filaments and the
sheets. The filaments observed in the galaxy distribution from the
SDSS has been shown to be statistically significant upto the length
scales of $80 \hmpc$ \citep{bharadwaj2004, pandey1}. Filaments are
elongated structures with a length of tens of Mpc \citep{colberg} and
thickness of $\sim 2-3 \hmpc$ \citep{gonzalez}. They can be of
different sizes and types (straight, warped, irregular etc.) based on
their visual morphology \citep{pimbblet}. The filaments are believed
to host $\sim 50 \%$ of the baryons in the Universe \citep{cen} and
are expected to play an important role in the formation and evolution
of galaxies. The filaments which are one of the most prominent visual
features in the galaxy distribution has so far drawn a lot of
attention in the literature. Contrary to this, the detection of sheets
or the walls in the galaxy distribution has attracted very little or
no attention at all.

There are also giant structures like the Sloan Great Wall
\citep{gott05} extending over length scales of more than 400 Mpc. The
Saraswati supercluster \citep{bagchi} which spans at least 200 Mpc is
a massive supercluster recently found in the SDSS. On the other hand,
the empty regions or the voids constitute of about $\sim 95\%$ volume
of the Universe \citep{kauffmann, piran, hoyle, platen}. The voids
seen in the galaxy distribution have different sizes such as Bootes
void with a radius of 62 Mpc \citep{kirshner} and the Eridanus
supervoid which extends upto $\sim$ 300 Mpc \citep{szapudi}. The
existence of these giant structures illustrate the variety and
richness of the environments for galaxy formation and evolution in the
cosmic web. Galaxy environments are primarily characterized by the
local density which is known to play a central role in the galaxy
formation and evolution. It has been also argued that besides the
density, the morphology of the environment may also play a crucial
role in the formation and evolution of galaxies \citep{pandey2,
  scudder, darvish, lupa, filho, pandey3, lee}. It would be
interesting to measure the relative abundance of these structures on
different length scales and understand their roles in the galaxy
formation and evolution.

The Sloan Digital Sky Survey (SDSS) which is currently the largest
redshift survey has mapped the distribution of millions of galaxies in
the nearby Universe providing an unprecedented view of the cosmic web
in the nearby Universe. This provides an unique opportunity to unravel
the cosmic web in greater detail than ever possible. \citet{sarkar}
propose the local dimension which is a simple measure to characterize
the environment in which a galaxy is embedded inside the cosmic
web. This has been applied earlier to the SDSS DR7 data by
\citet{sarkar1} to study the length scale dependence and density
dependence of the various morphological components of the cosmic
web. The local dimension can be also employed to address several other
important issues related to the cosmic web. In this work, we analyze
the data from the SDSS DR14 with the local dimension to study how the
fraction of galaxies residing in different morphological environments
changes with the associated length scales. This allows us to explore
the relative abundance of different types of structures at different
length scales and identify the length scales which are dominated by
any particular type of structures. We also prepare a list of galaxies
for which the local dimension can be computed throughout the entire
length scale range available for this analysis. This would enable us
to track the gradual transition of the environment of a galaxy with
the increasing length scales.

We compare our findings with a semi analytic model of galaxy formation
by using a semi analytic galaxy catalogue \citep{henriques15} based on
the Millennium Run Simulation (MRS) \citep{springel}. Further, some of
the filaments and sheets observed in the galaxy distributions are the
outcome of random chance alignments. So we also compare our findings
against the random mock catalogues from Poisson distributions to
quantify the fraction of galaxies identified as part of filaments and
sheets which are the products of random chance alignment.

We also test the possible roles of any systematics such as the number
density and the cut-off in the goodness of fit in influencing the
results of the present analysis. Finally, we test the efficiency of
the method by applying it to a set of simulations of the segment Cox
process.

We convert redshifts to distances using a $\Lambda$CDM cosmological
model with $\Omega_{m0}=0.31$, $\Omega_{\Lambda0}=0.69$ and $h=1$
throughout the analysis. 

A brief outline of our paper is as follows. We describe the method in
Section 2 and the data in Section 3. We present our results
and conclusions in section 4 and section 5 respectively.

\begin{table*}
\caption{This table shows the criteria to classify the galaxies into
  different classes based on their local dimension.}
\label{tab:gclass}
\begin{tabular}{|r|c|c|c|c|c|}
\hline
Class: & $C1$ & $C2$ & $C3$ & $I1$ & $I2$ \\
\hline
Local dimension : &   $ 0.75 \le D < 1.25 $  &     $ 1.75 \le D < 2.25$   &   $ D \ge 2.75 $    &    $ 1.25 \le D < 1.75$    &   $ 2.25 \le D < 2.75 $  \\
\hline
\end{tabular}
\end{table*}

\section{METHOD OF ANALYSIS}

We consider a sphere of radius $R$ centred around each galaxy in the
volume limited sample. The centres for which the spheres remain
completely inside the survey boundary are identified and we count the
number of galaxies $N(<R)$ inside each of these spheres. We repeat
these measurements for a number of different radius $R$ within a
specified length scale range $R_1\leq R \leq R_2$. The value of $R_1$
is kept fixed and $R_2$ is gradually increased upto the largest radius
accessible within the survey region.

The cosmic web is an interconnected network of sheets, filaments and
clusters and each of the galaxies are part of any of these structural
elements. We expect the number of galaxies within a sphere of radius
$R$ centered around a galaxy to scale as,
\begin{eqnarray}
 N(< R)= A R^{D}
\label{eq:ld1}
\end{eqnarray}
where A is a constant and the exponent $D$ is the local dimension
\citep{sarkar}. The local dimension $D$ quantifies the nature of the
structural element in which it is embedded. We expect $D=1$ and $D=2$
for galaxies residing in the filaments and sheets respectively. $D=3$
around a galaxy can both indicate a galaxy cluster or any volume
filling structures such as the cosmic web on large scales. It may be
noted that the intermediate values of the local dimension $D$ are also
possible when the counting sphere incorporates multiple structural
elements of different types.

We fit the galaxy counts $N(<R)$ around each centre to
\autoref{eq:ld1} and measure the $D$ value associated with each
galaxy. The local neighbourhood of a galaxy is expected to look
different at different length scales. Consequently, the measured $D$
values are expected to change with the increasing length scales and
finally approach $D\sim3$ when the galaxy is surrounded by a
homogeneous network. This would occur only beyond the scale of
homogeneity. 

\begin{figure*} 
\resizebox{8.5 cm}{!}{\rotatebox{0}{\includegraphics{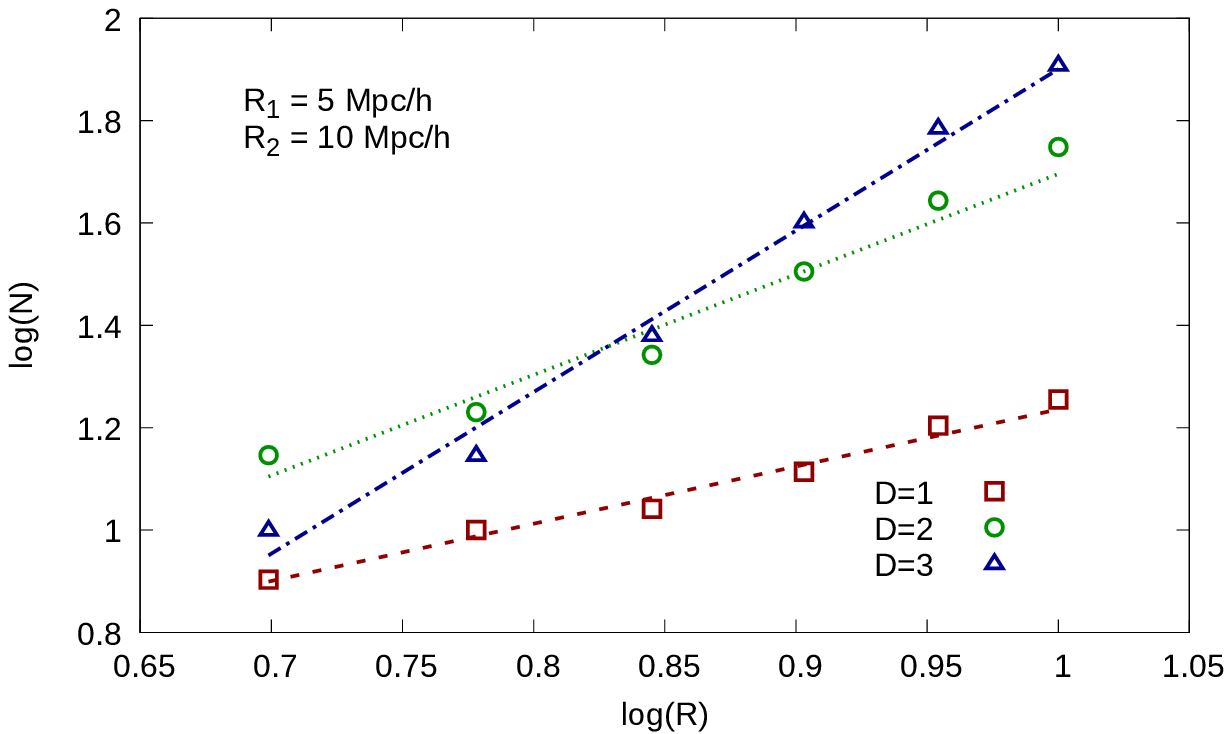}}} %
\hspace{0.5 cm}
\resizebox{8.5 cm}{!}{\rotatebox{0}{\includegraphics{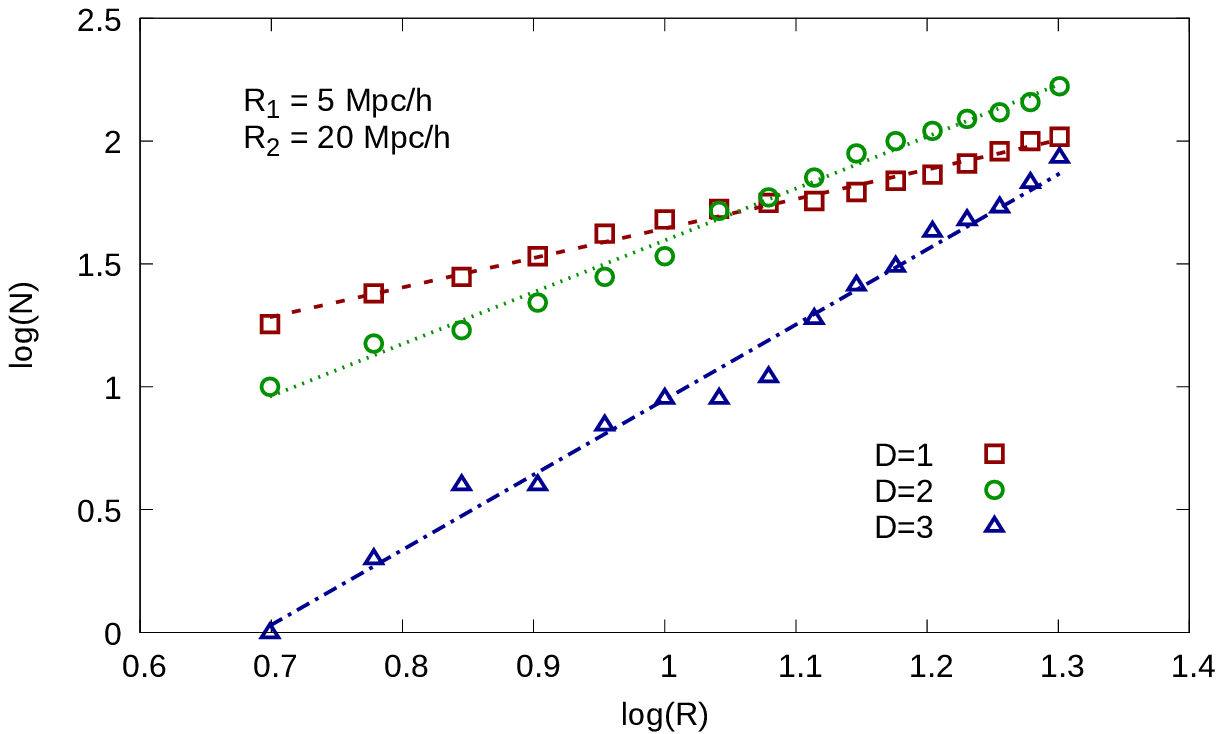}}}
\caption{The left panel shows the best fit lines along with the
  measured values of $N$ as a function of $R$ for three different
  galaxies with local dimension 1,2 and 3 respectively. The fits are
  carried within the length scale range $5 \hmpc \leq R \leq 10 \hmpc$
  and each of these fits satisfies the criteria $\chi ^2 / \nu \le
  \frac{1}{2} $ employed in this work. The right panel shows the same
  for another three galaxies with $D=1,2,3$ for which the number
  counts are fitted within length scale range $5 \hmpc \leq R \leq 20
  \hmpc$.}
  \label{fig:Dfit}
\end{figure*}

We consider only those centres for which we have at least 10
neighbouring galaxies within radius $R_2$. The value of $D$ for each
galaxy within each length scale range $R_1 \le R \le R_2$ is estimated
using a least-square fit and a $\chi^2$ value is also calculated for
each of these fits. We apply a cut in the Chi-square per degree of
freedom $\frac{{\chi}^2}{\nu} \le 0.5 $ to identify only the good
quality fits for our analysis (\autoref{fig:Dfit}). We classify the
galaxies into five classes based on the measured values of their local
dimension $D$. \autoref{tab:gclass} provides the criteria for this
classification. C1 and C2 are the galaxies which are part of a
filament or sheet respectively. The C3 galaxies are part of volume
filling structures. The I1 and I2 galaxies with intermediate $D$
values may lie near the junction of two different types of structural
elements.

For each length scale range $R_1\leq R \leq R_2$, we find the number
and fractions of classified galaxies in each class.

\begin{table*}
\caption{This table shows the parameters used to simulate the datasets
  for the segment Cox process.}
\label{tab:scox}
\begin{tabular}{|c|c|c|c|c|}
\hline
Length of & Mean number of points per unit length & Total number of\\
 segments &  on the segments & segments\\
 $(\hmpc)$&$(h\,\rm{Mpc^{-1}})$&\\
\hline
10   & 1 & 3000   \\
30   & 1 & 500    \\
50   & 1 & 100    \\ 
\hline
\end{tabular}
\end{table*}

\section{DATA}
\subsection{SDSS DR14}
 
We use data from the $14^{th}$ data release of the Sloan Digital Sky
Survey (SDSS) \citep{abolfathi17} which is the second data release of
the fourth phase (SDSS IV) of the survey. DR14 has accumulated
spectral and imaging data taken from August 2014 to July 2016 by the SDSS
2.5 m telescope and it has the most current and reprocessed data that
incorporates the entire coverage of the prior data releases. We use a
Structured Query Language (SQL) to get the data from SDSS
CasJobs\footnote{http://skyserver.sdss.org/casjobs/}. We select a
contiguous region in the Northern galactic hemisphere using the cuts $
0^{\circ} \leq \delta \leq 60^{\circ}$ and $ 135^{\circ} \leq \alpha
\leq 225^{\circ}$ , where $\alpha$ and $\delta$ are the right
ascension and declination respectively. We select all the galaxies
within redshift $z<0.3$ and r-band Petrosian magnitude $m_r<17.77$ in
this region. We set the $ZWARNING$ flag to zero to select only the
galaxies with good spectrum and reliable redshift. These cuts yield a
total $377606$ galaxies. We then prepare a volume limited sample from
this data by applying a cut $M_r<-20.5$ in the K-corrected and
extinction corrected $r$-band absolute magnitude. The K-corrections
are obtained from a polynomial fit provided by \citet{park05}. The
resulting volume limited sample contains $90406$ galaxies within
redshift $z<0.1385$ which radially extends upto $406 \hmpc$. The
galaxy sample has a number density of $\sim 2.977 \times 10^{-3} h^{3}
Mpc^{-3}$ and the mean intergalactic separation of $\sim 6.95 \hmpc$.

\subsection{Millennium Run Simulation}
The Millennium Run Simulation (MRS) \citep{springel} is one of the
largest high resolution cosmological N-body simulation available
todate. The Millennium simulation followed the evolution of $2160^3$
dark matter particles in a comoving box of size $500 \hmpc$ from
redshift $z=127$ to $z=0$. The semi analytic models (SAM)
\citep{white91,kauff1,cole1,bagh,somervil,benson} provide a powerful
and effective tool to study the galaxy formation and evolution. These
models parametrise the physics involved in terms of simple models
following the dark matter merger trees over time and finally provide
the statistical predictions of galaxy properties at any desired
epoch. Here we use the data from a semi analytic galaxy catalogue
\citep{henriques15} derived from the Millennium run simulation
\citep{springel}. \citet{henriques15} updated the Munich model of
galaxy formation using the values of cosmological parameters from
PLANCK first year data. We use a SQL query to extract the data from
the Millennium
database \footnote{https://wwwmpa.mpa-garching.mpg.de/millennium/}.
We map the Millennium galaxies to redshift space by using their
peculiar velocities and then construct the mock samples by applying
the same absolute magnitude cut as applied to the SDSS data. We ensure
that the mock sample has the identical geometry and number density as
the actual SDSS sample. We construct $10$ such mock SDSS samples from
the SAM catalogue from the Millennium Run simulation by placing the
observer at different location. These mock samples are not derived
from independent regions as we have only one realization of the SAM
catalogue.

\subsection{Poisson sample}

We construct 10 mock SDSS samples from Poisson distributions. These
mock random catalogues have exactly the same geometry and number
density as the actual SDSS sample used in this analysis.

\subsection{Segment Cox process}

We simulate a set of segment Cox process \citep{martinez,borderia}
inside a cube of sides $250 \hmpc$ to test the efficiency of the
method employed in the present work. The segment Cox process is a
controlled point process where segments of length $l$ are scattered
with random positions and orientations over a given volume. We first
generate a random position and then choose a random orientation for a
segment. The segment is then populated with points at random locations
on it. The process is repeated for the desired number of segments. The
segment length, the number of segments per unit volume and the mean
number of points per unit length of the segments are the control
parameters in the segment Cox process. We generate $10$ realizations
of the segment Cox process each with segment length $10 \hmpc$, $30
\hmpc$, and $50 \hmpc$. The control parameters of the simulated
datasets are described in \autoref{tab:scox}.

\section{RESULTS}

\begin{figure*} 
\resizebox{7 cm}{!}{\rotatebox{0}{\includegraphics{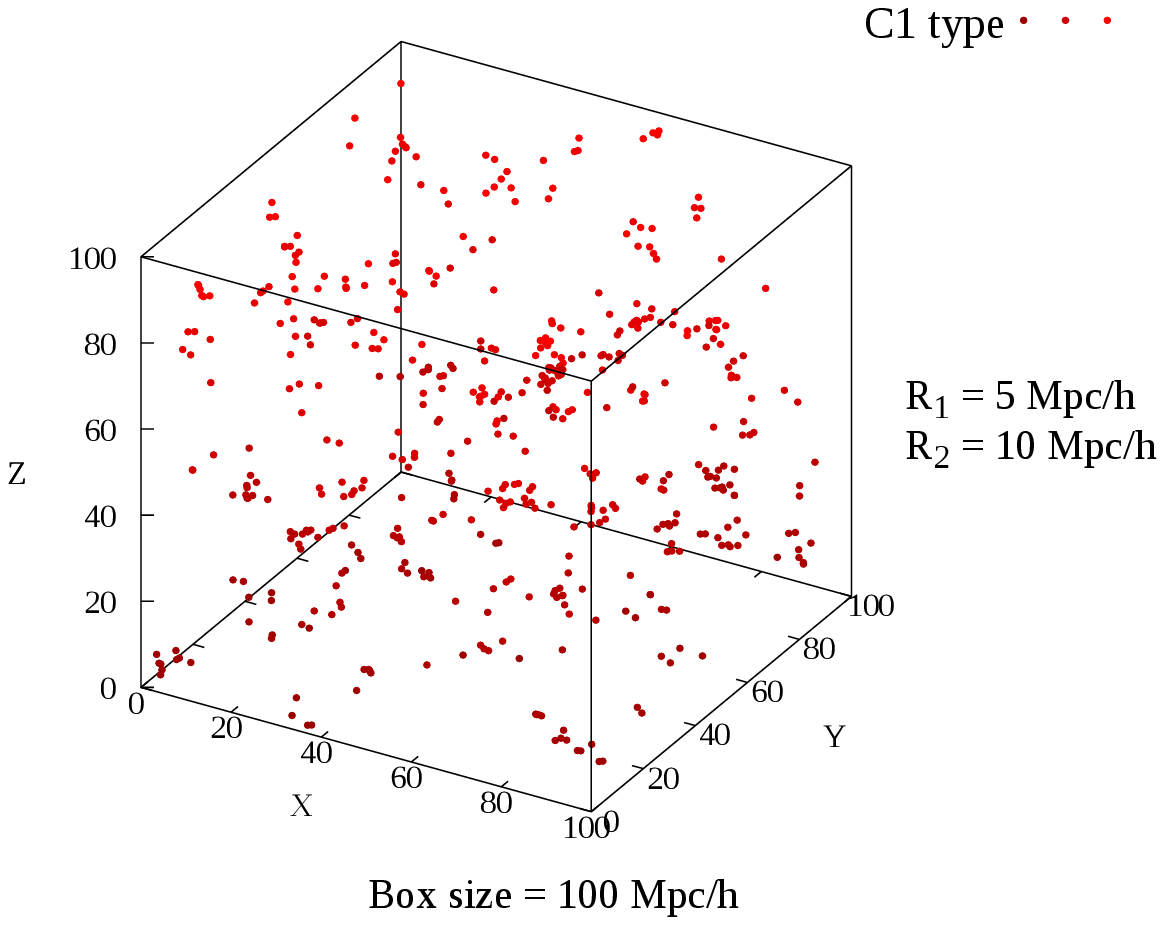}}} %
\hspace{2 cm}
\resizebox{7 cm}{!}{\rotatebox{0}{\includegraphics{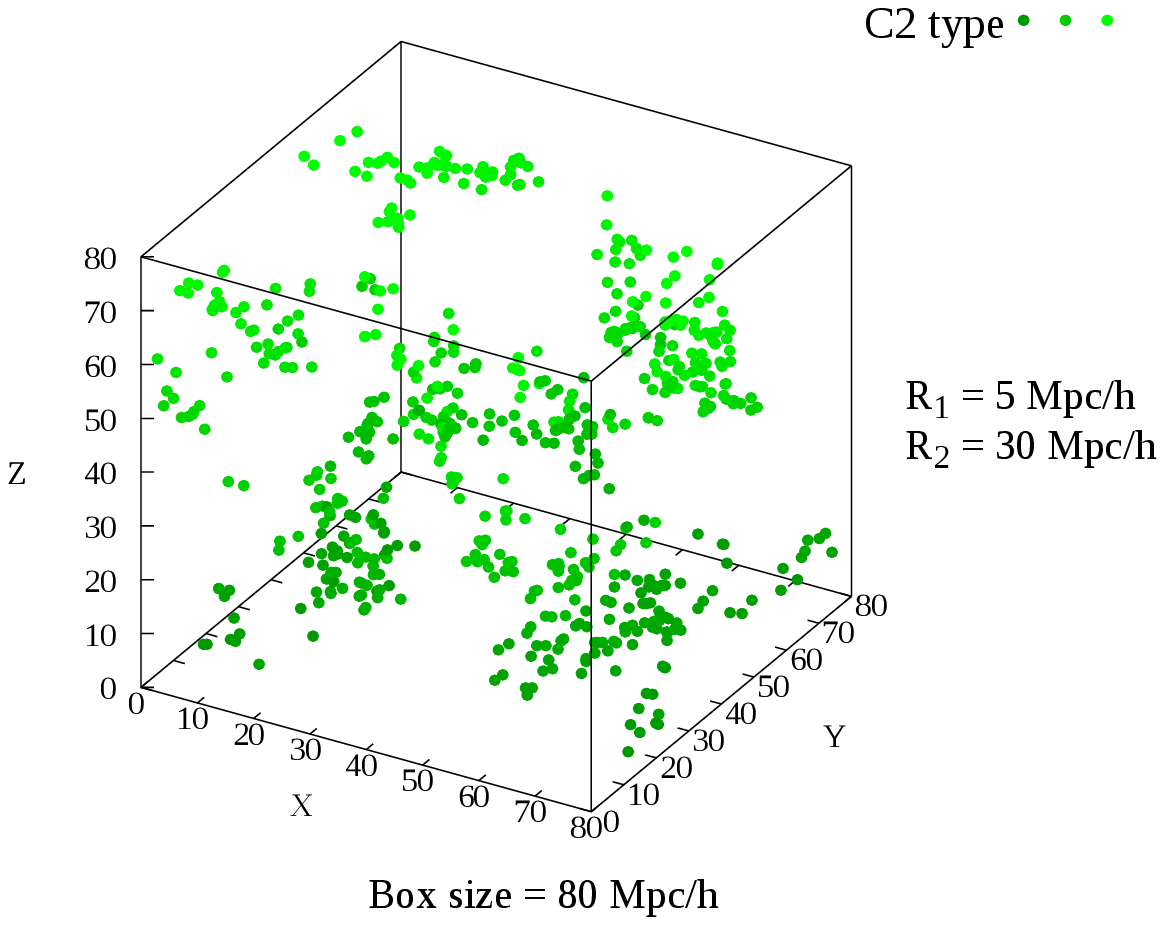}}} %
\vspace{0.5 cm}
\resizebox{7 cm}{!}{\rotatebox{0}{\includegraphics{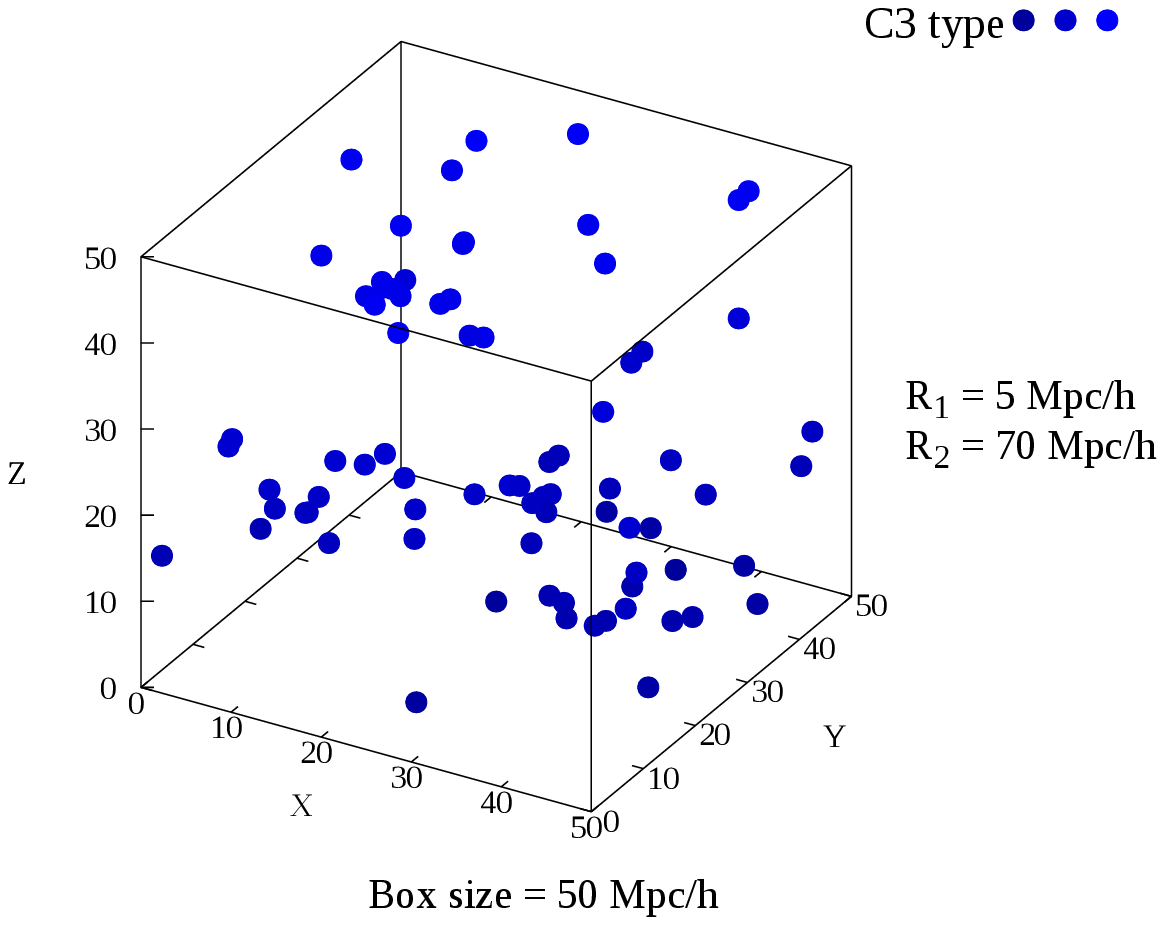}}}
\caption{ The top left box shows the distribution of $C1$ type
  galaxies in a cubic region of side $ 100 \hmpc$. The cubic region
  represents only a part of the SDSS sample analyzed and the
  co-ordinates shown here are not the real co-ordinates but the
  co-ordinates scaled according the geometry of the box. The top right
  box shows the distribution of $C2$ type galaxies in a cube of side
  $80 \hmpc$ and the bottom middle box shows the distribution of $C3$
  type galaxies in a cube of side $50 \hmpc$. The values of $R_1$ and
  $R_2$ in each panel indicate the length scales over which the number
  counts are fitted to compute the local dimension.}
  \label{fig:3Dview_ow}
\end{figure*}

\begin{figure*}
\resizebox{8.5 cm}{!}{\rotatebox{0}{\includegraphics{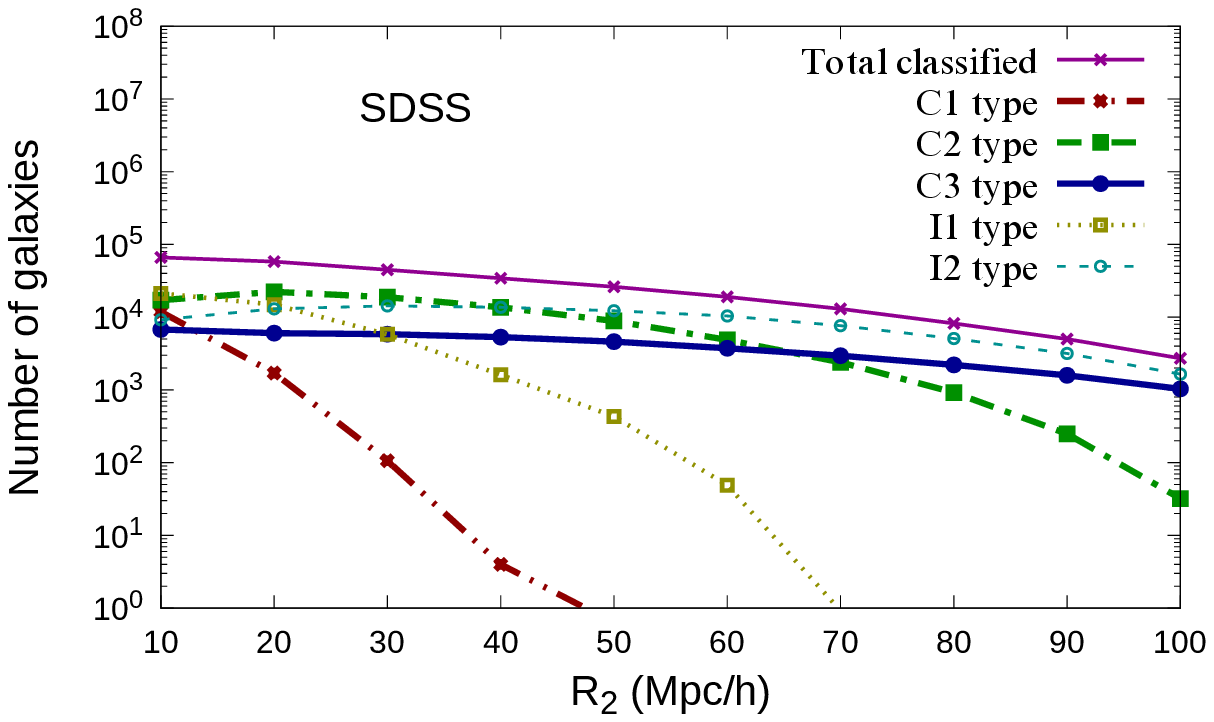}}} %
\hspace{0.5 cm}
\resizebox{8.5 cm}{!}{\rotatebox{0}{\includegraphics{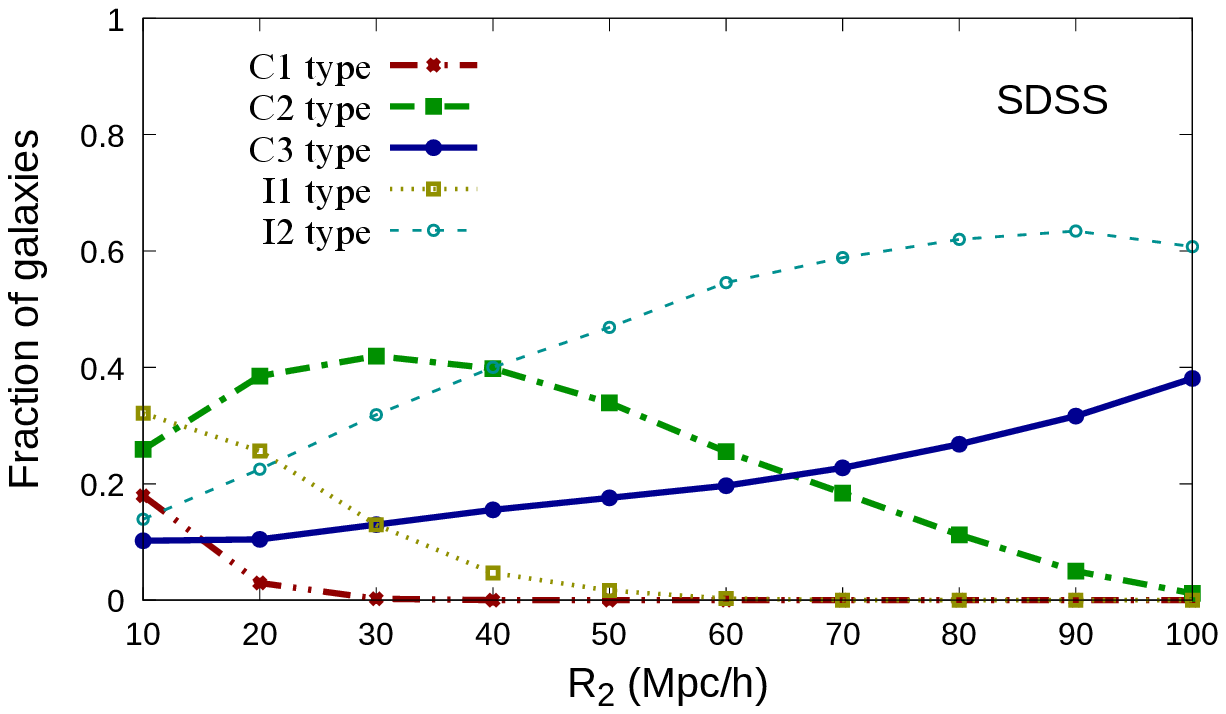}}}
\resizebox{8.5 cm}{!}{\rotatebox{0}{\includegraphics{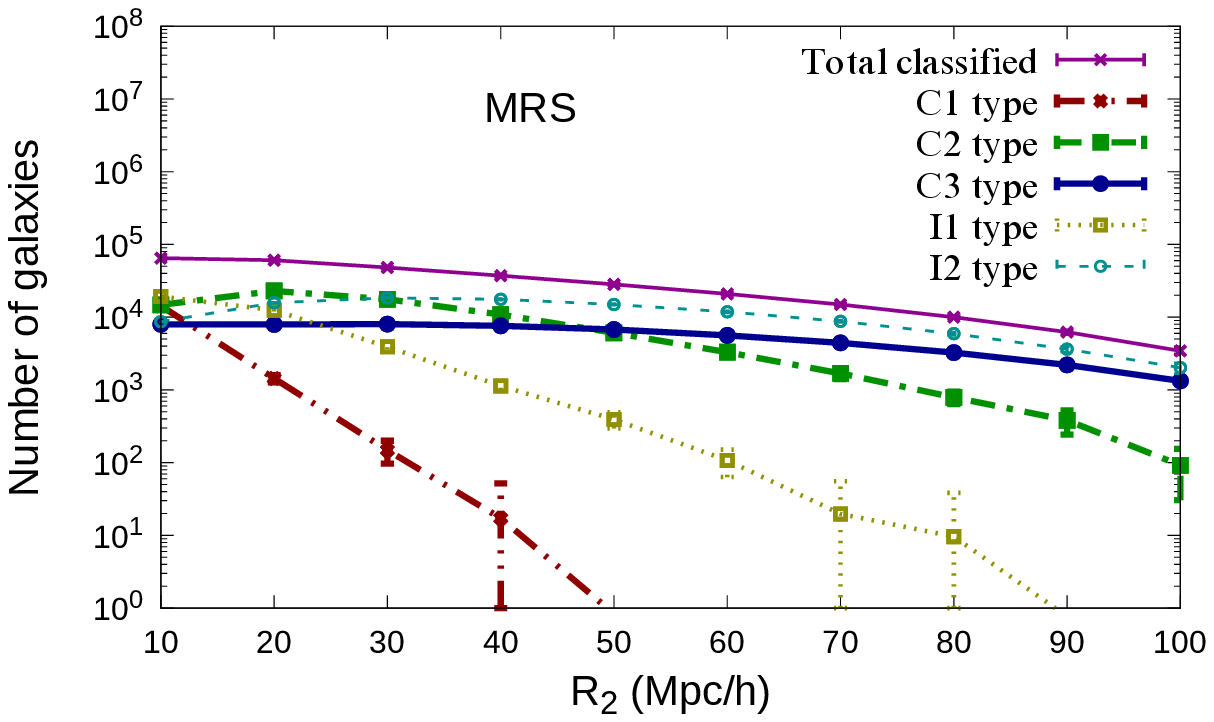}}} %
\hspace{0.5 cm}
\resizebox{8.5 cm}{!}{\rotatebox{0}{\includegraphics{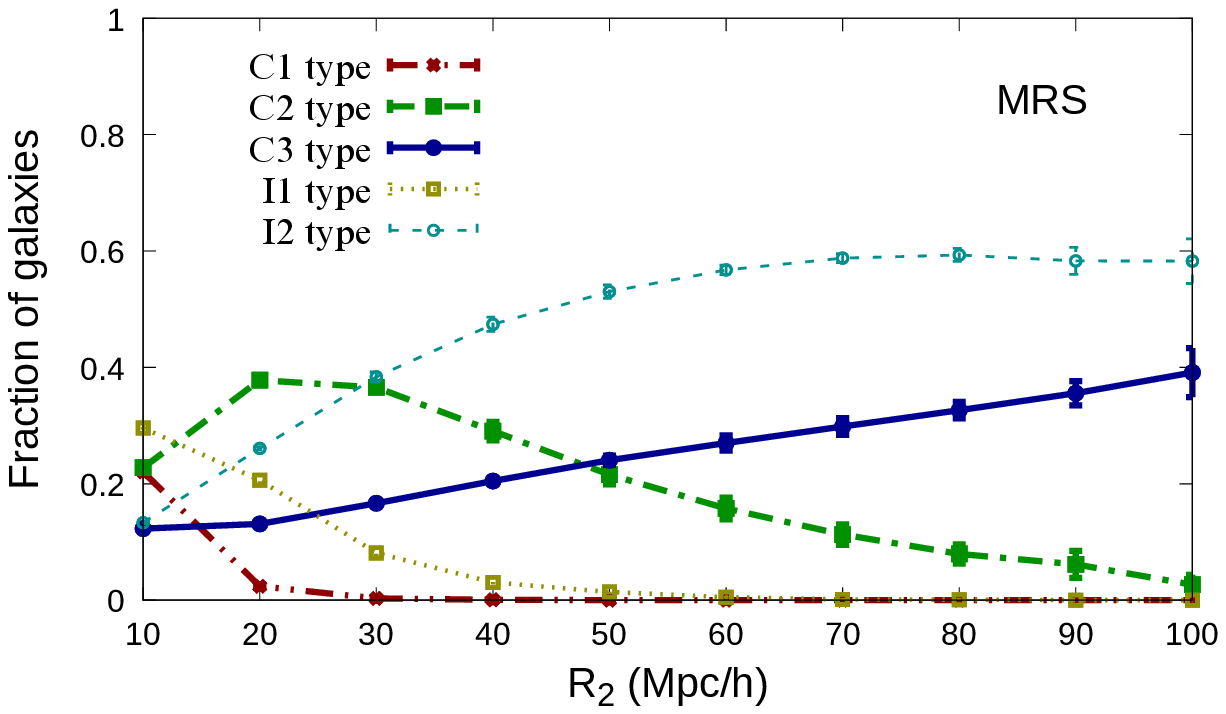}}}
\resizebox{8.5 cm}{!}{\rotatebox{0}{\includegraphics{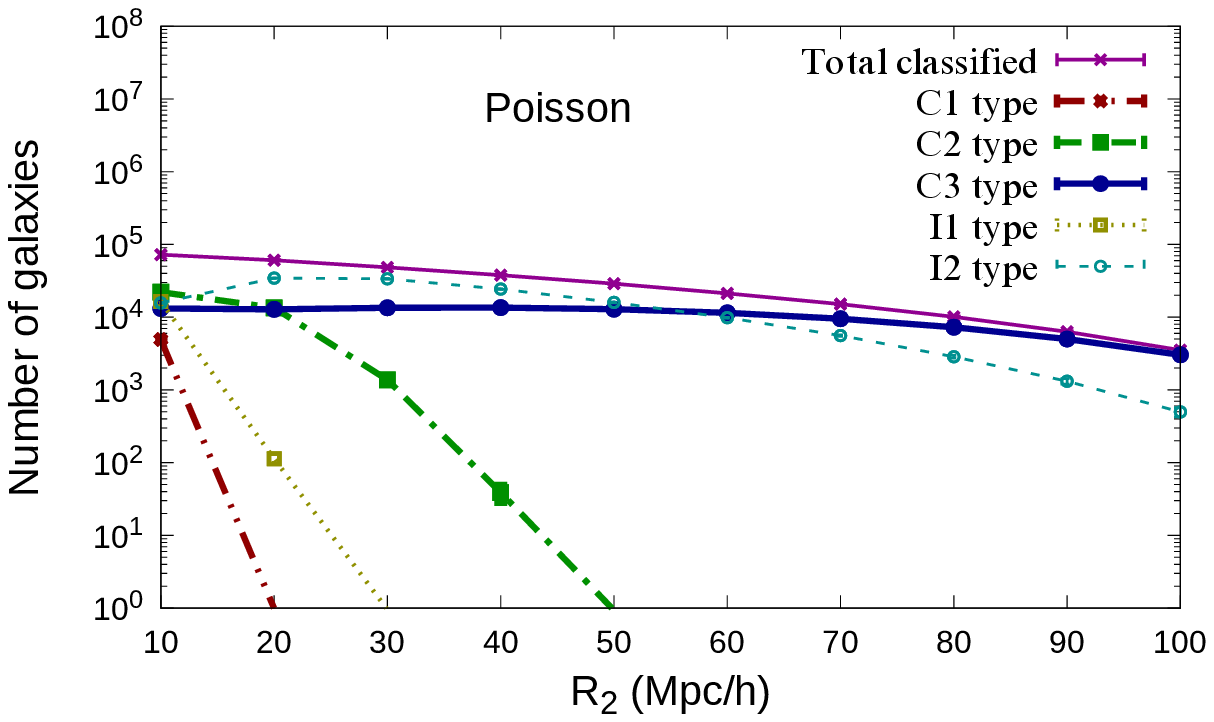}}} %
\hspace{0.5 cm}
\resizebox{8.5 cm}{!}{\rotatebox{0}{\includegraphics{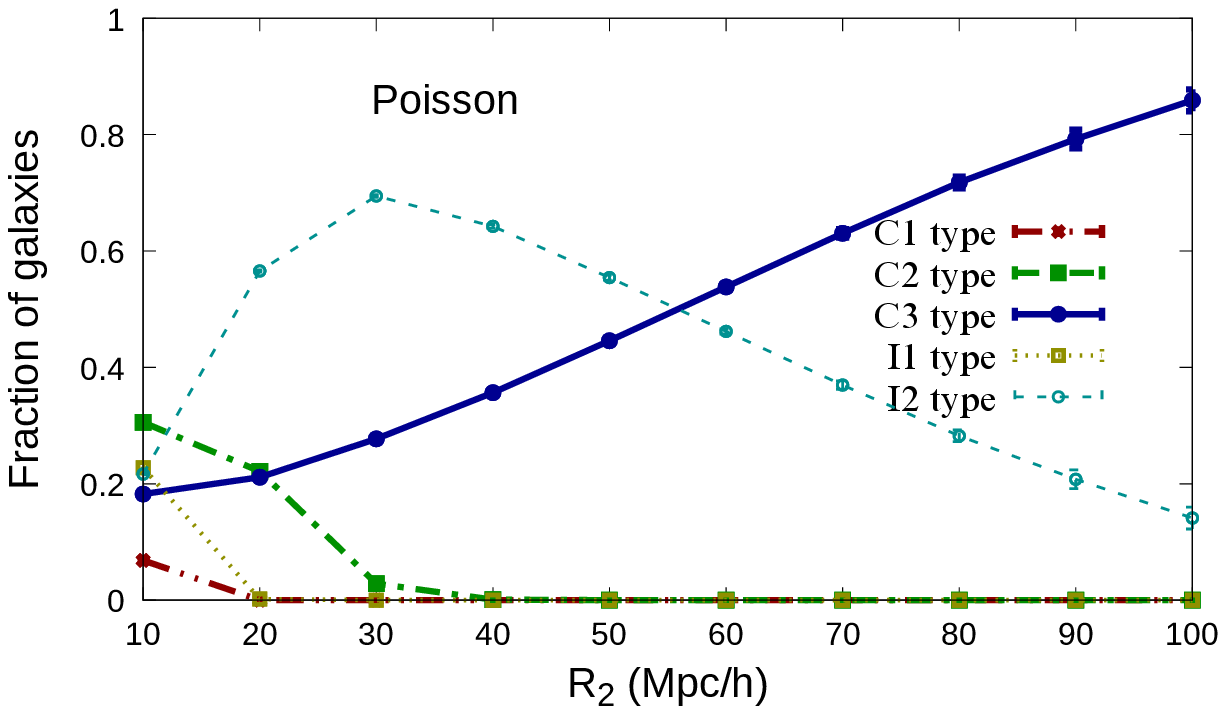}}}
\caption{The top left panel shows the number of different types of
  SDSS galaxies classified according to their local dimension
  (\autoref{tab:gclass}) at different values of $R_2$. The right panel
  shows how the fraction of different types of galaxies vary with
  increasing value of $R_2$. The two middle panels and the two bottom
  panels show the same but for the mock galaxy samples from a semi
  analytic galaxy catalogue from the Millennium simulation and the
  Poisson distributions respectively. The value of $R_1$ is fixed at
  $5 \hmpc$ in each case. The error-bars shown for the Poisson
  distributions and Millennium simulation are obtained from $10$
  independent realizations. The size of the error-bars are very small
  for the Poisson distributions. The error-bars are also very small
  for the Millennium simulation as all the $10$ mock samples are
  derived from the same catalogue.}
  \label{fig:nclass_fclass}
\end{figure*}

\begin{figure*}
\resizebox{8 cm}{!}{\rotatebox{0}{\includegraphics{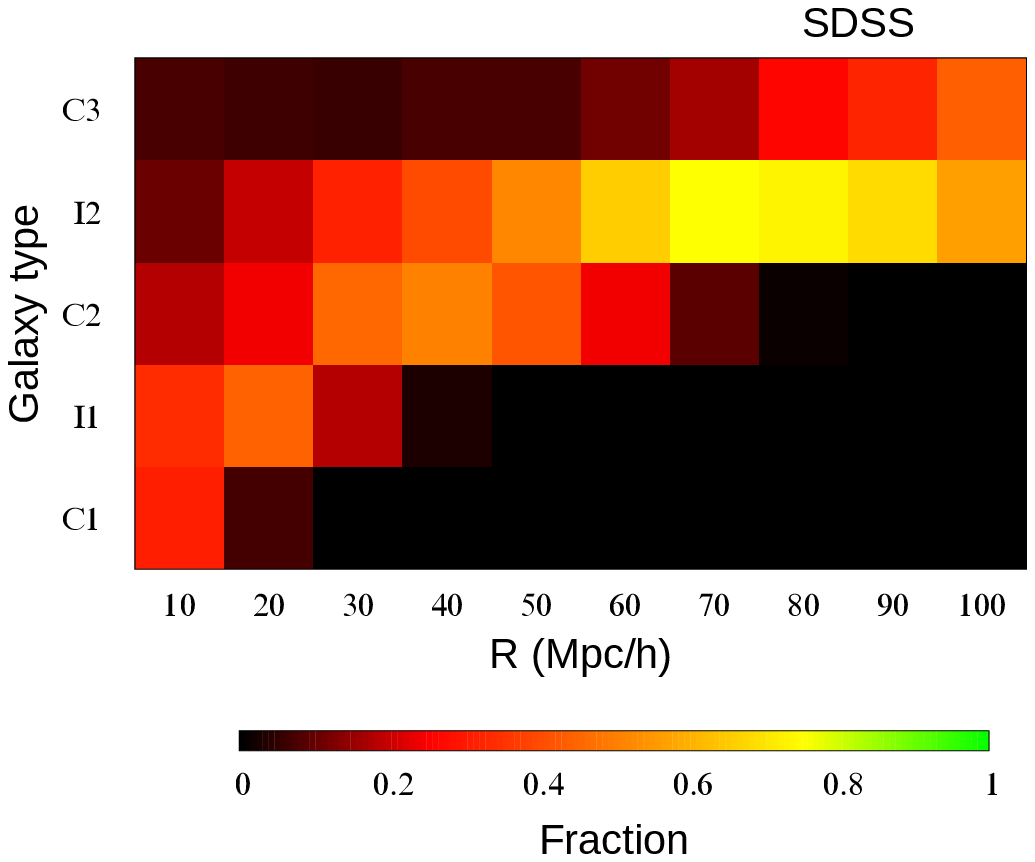}}}%
\hspace{0.5 cm}
\resizebox{8 cm}{!}{\rotatebox{0}{\includegraphics{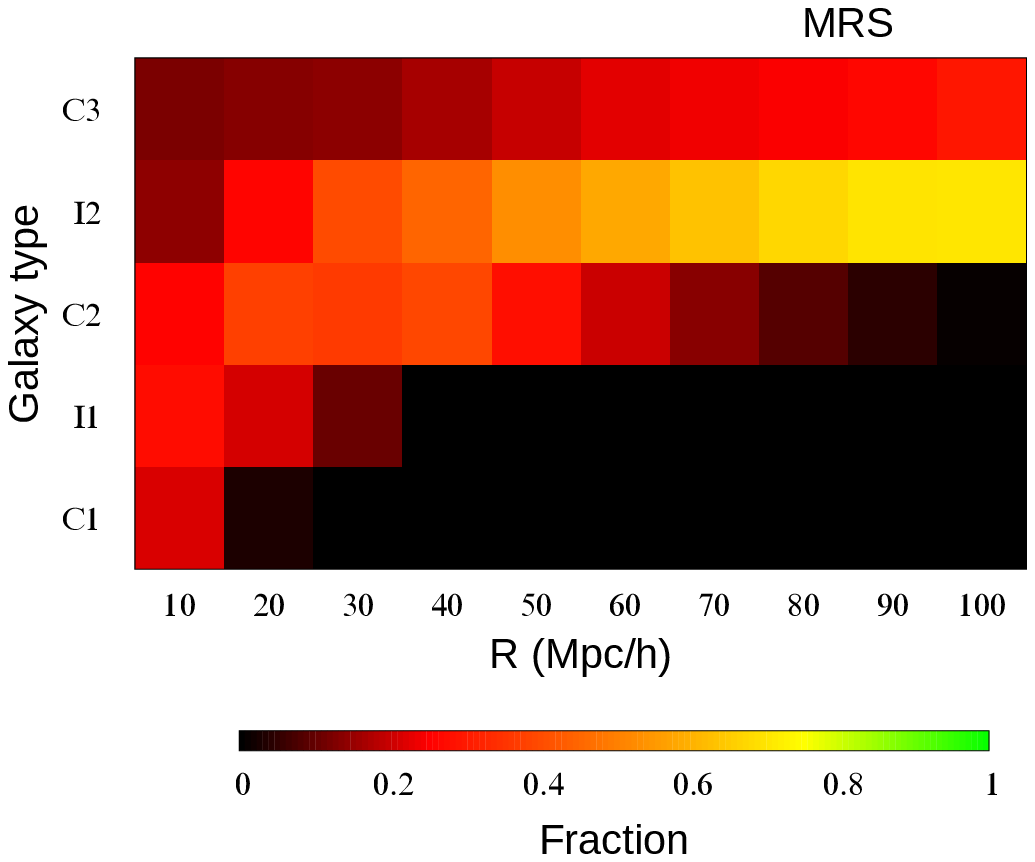}}} %
\hspace{0.5 cm}
\resizebox{8 cm}{!}{\rotatebox{0}{\includegraphics{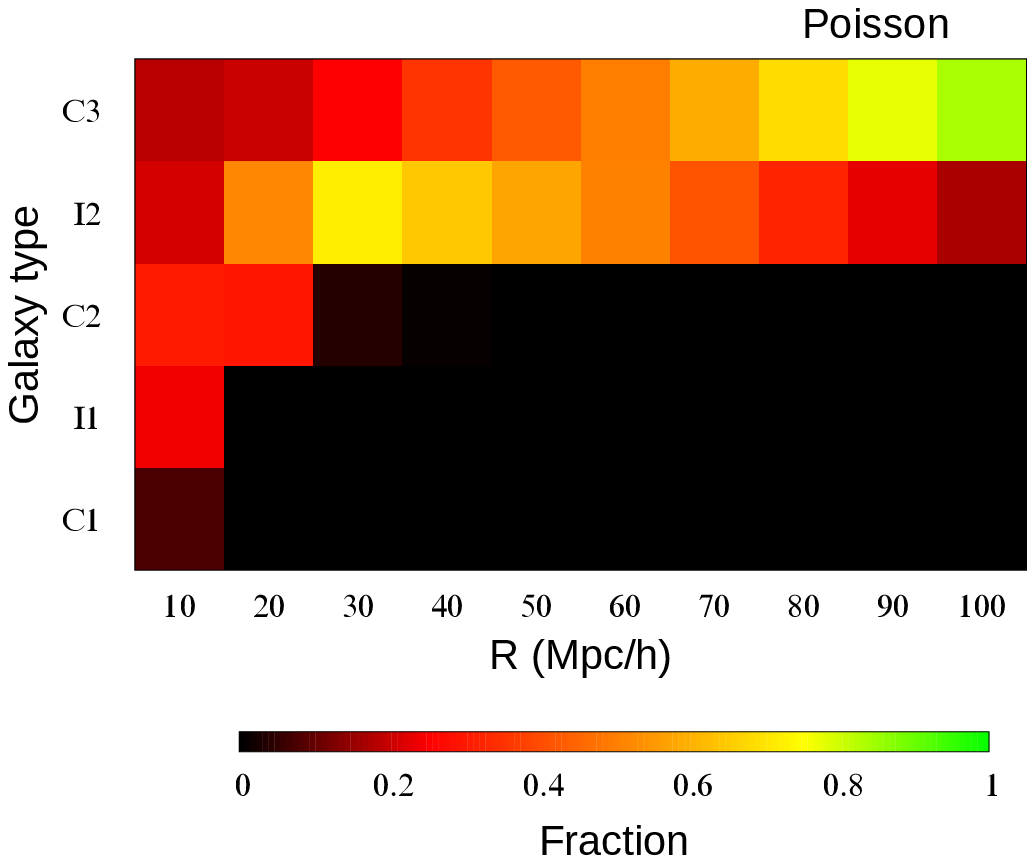}}}
\caption{ The top left panel shows the fraction of different types of
  SDSS galaxies at different length scales, only for the galaxies for
  which the local dimension can be computed at each and every length
  scale. The top right and the bottom middle panel show the same for
  the mock galaxy samples from a semi analytic galaxy catalogue from
  the Millennium simulation and the Poisson distributions
  respectively.}
  \label{fig:Dtrans_all}
\end{figure*}

\begin{figure*}
\resizebox{8.5 cm}{!}{\rotatebox{0}{\includegraphics{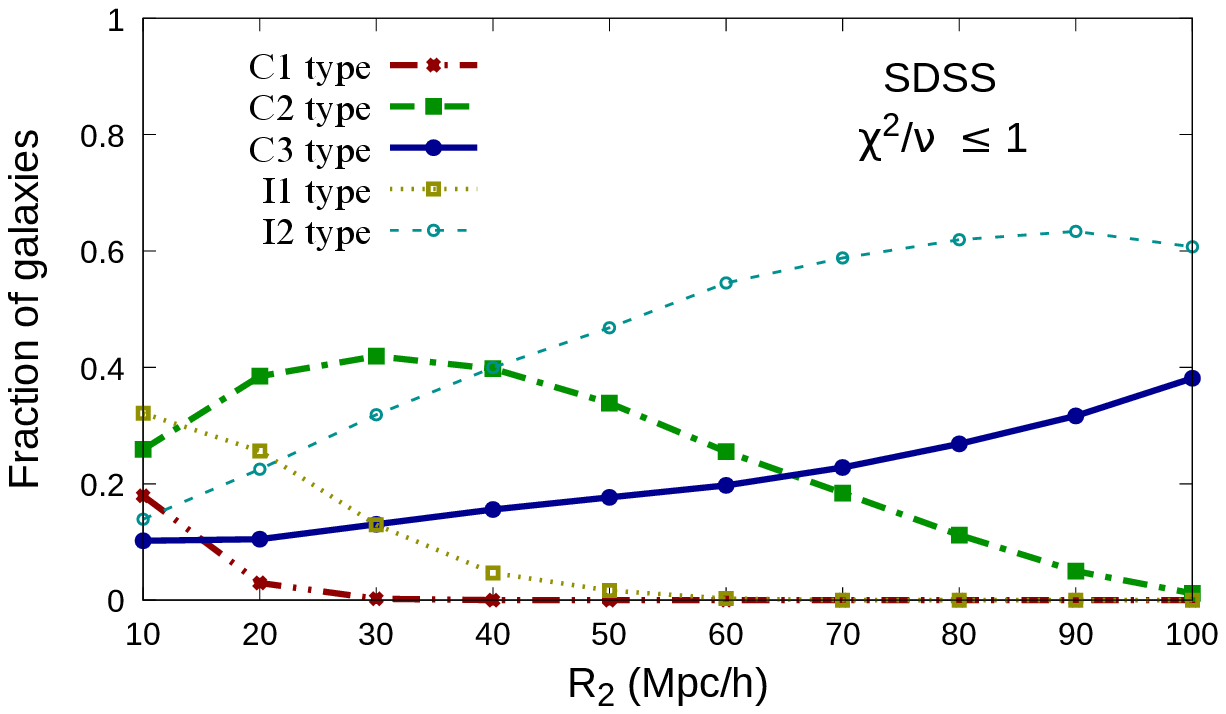}}} %
\resizebox{8.5 cm}{!}{\rotatebox{0}{\includegraphics{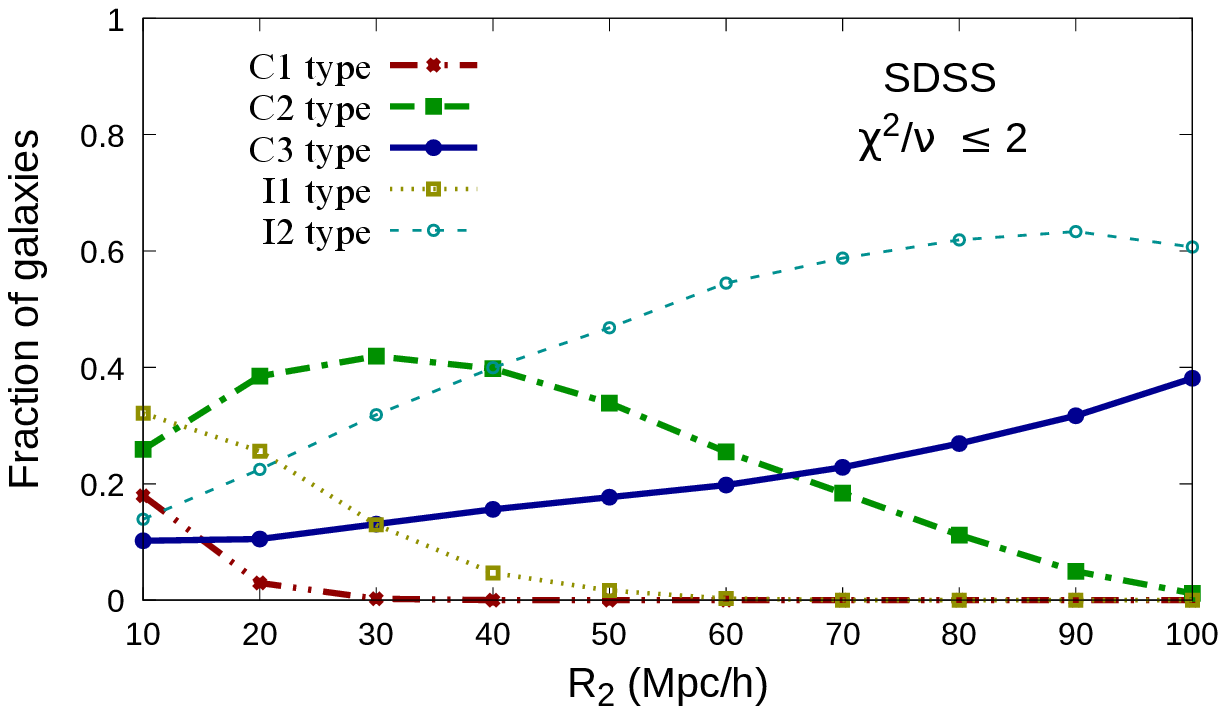}}}\\
\resizebox{8.5 cm}{!}{\rotatebox{0}{\includegraphics{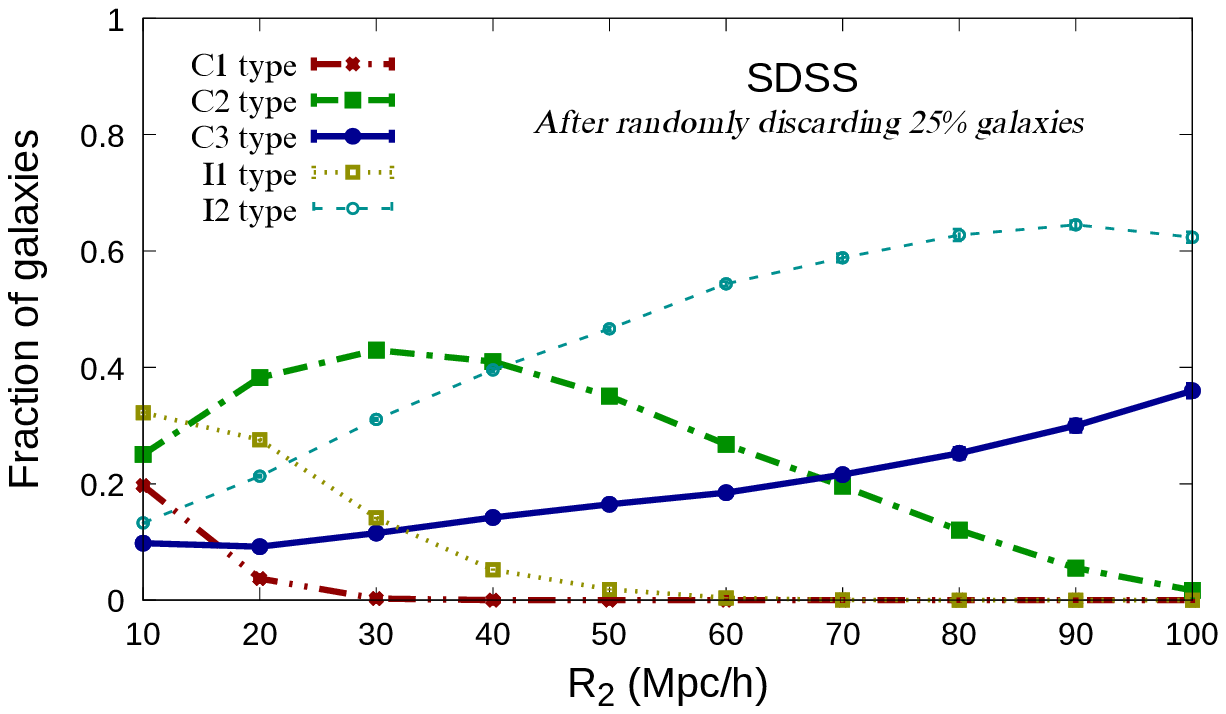}}} %
\resizebox{8.5 cm}{!}{\rotatebox{0}{\includegraphics{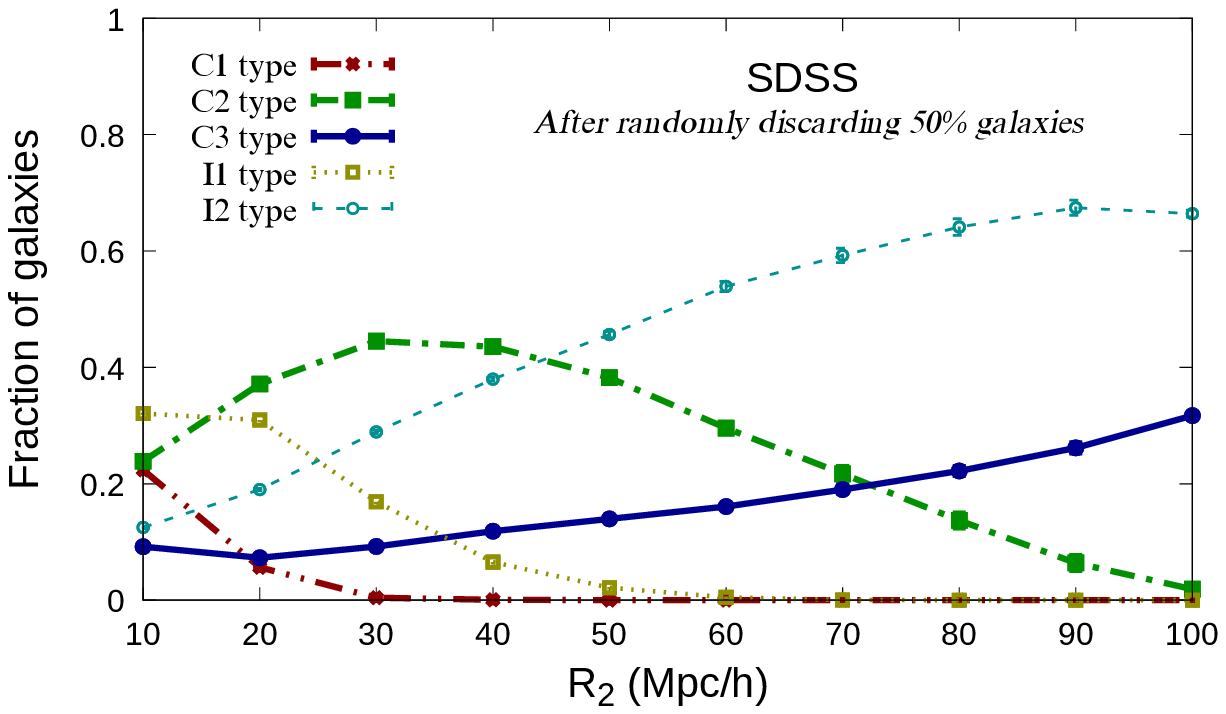}}}\\
\resizebox{8.5 cm}{!}{\rotatebox{0}{\includegraphics{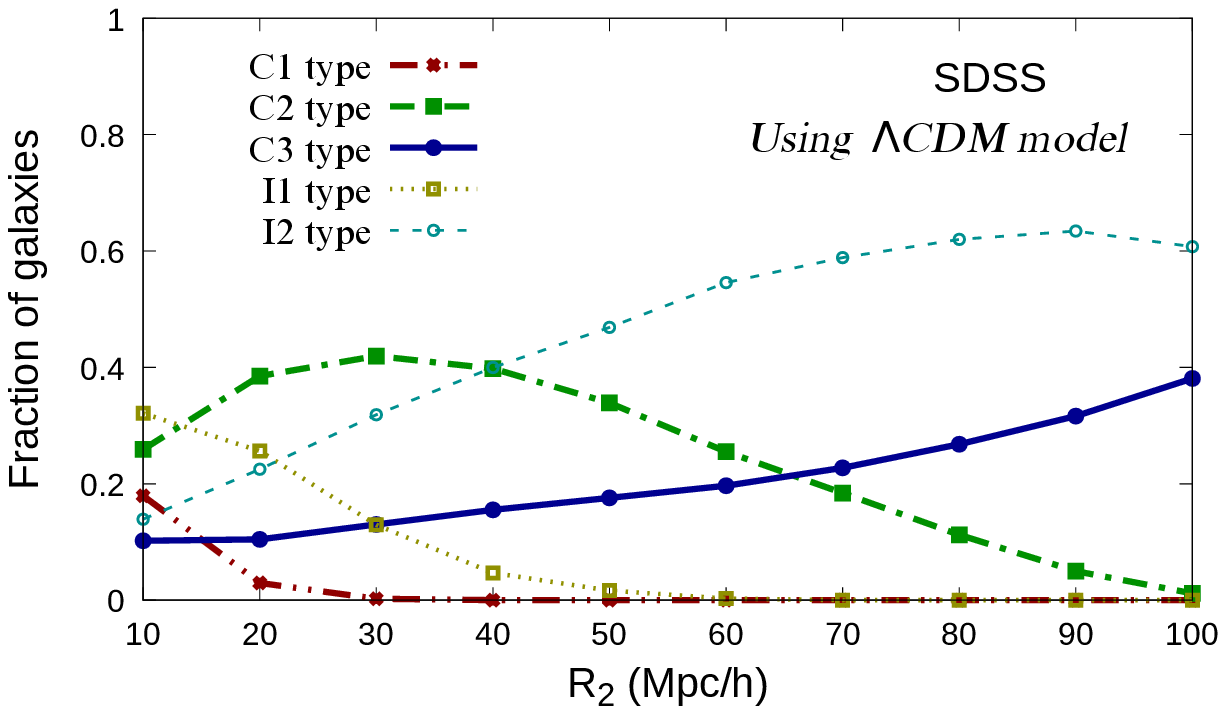}}} 
\resizebox{8.5 cm}{!}{\rotatebox{0}{\includegraphics{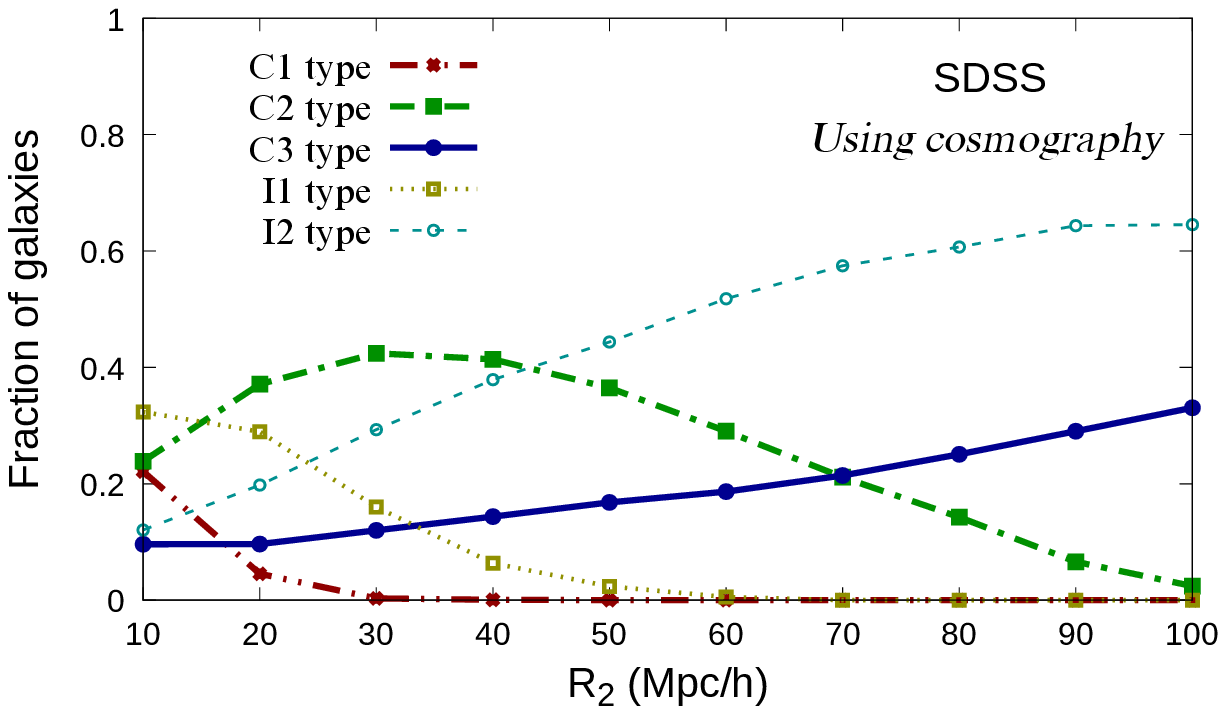}}} %
\caption{The top left and right panels of the figure show the fraction
  of different types of galaxies as a function of length scale when
  the good quality fits are selected using the cut-offs
  $\frac{\chi^2}{\nu} \leq 1$ and $\frac{\chi^2}{\nu} \leq 2$
  respectively. The middle left and right panels respectively show the
  results with cut-off $\frac{\chi^2}{\nu} \leq 0.5$ but after
  randomly discarding $25\%$ and $50\%$ galaxies from the original
  volume limited sample. The two bottom panels show the results for
  the SDSS volume limited sample when the redshifts are converted to
  distances using the $\Lambda$CDM model and cosmography. The
  1-$\sigma$ errorbars are only shown in the two middle panels which
  are drawn from 10 different subsamples.}
  \label{fig:systematic}
\end{figure*}

\begin{figure*}
\resizebox{7.5 cm}{!}{\rotatebox{0}{\includegraphics{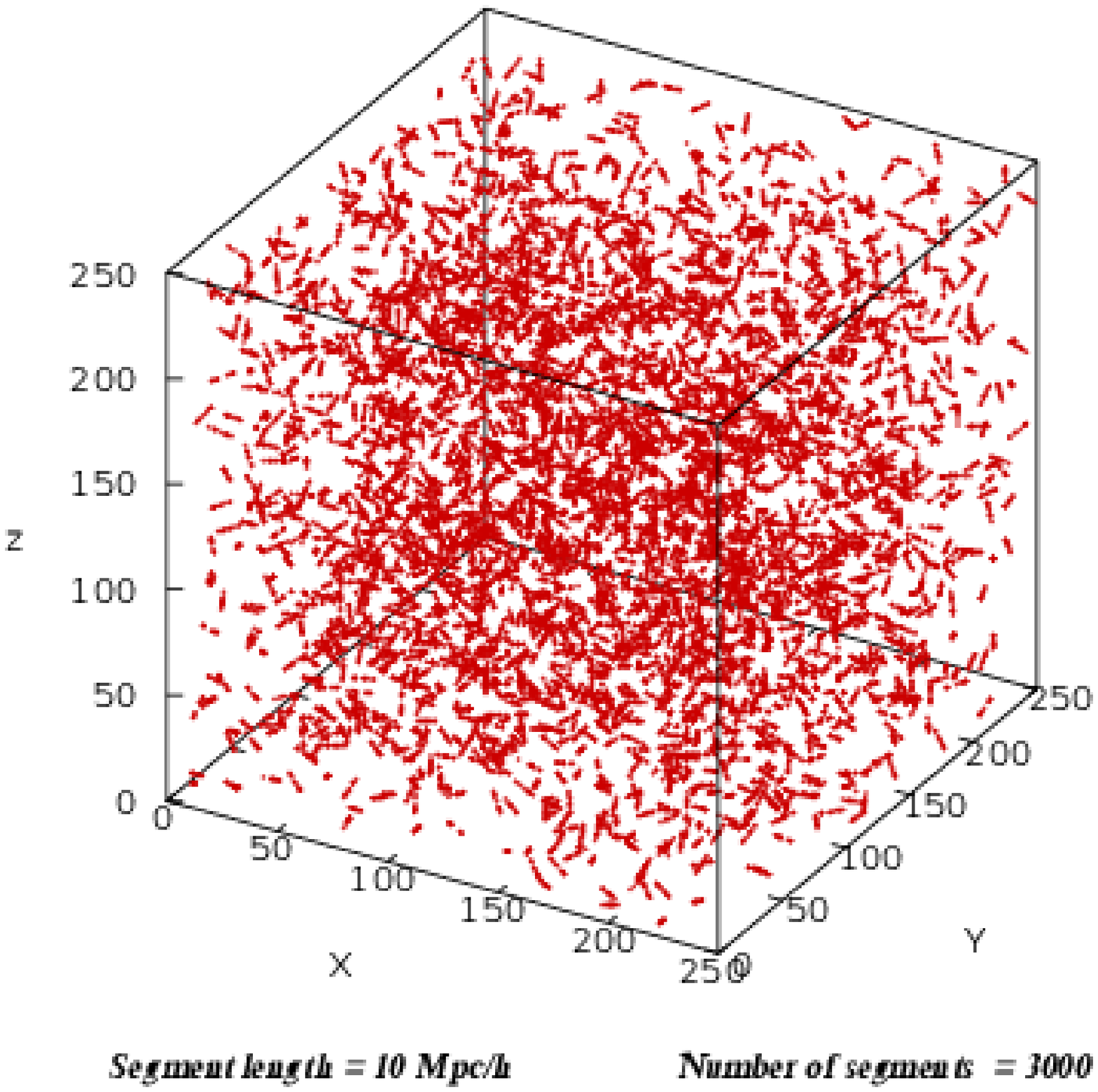}}} %
\resizebox{8.5 cm}{!}{\rotatebox{0}{\includegraphics{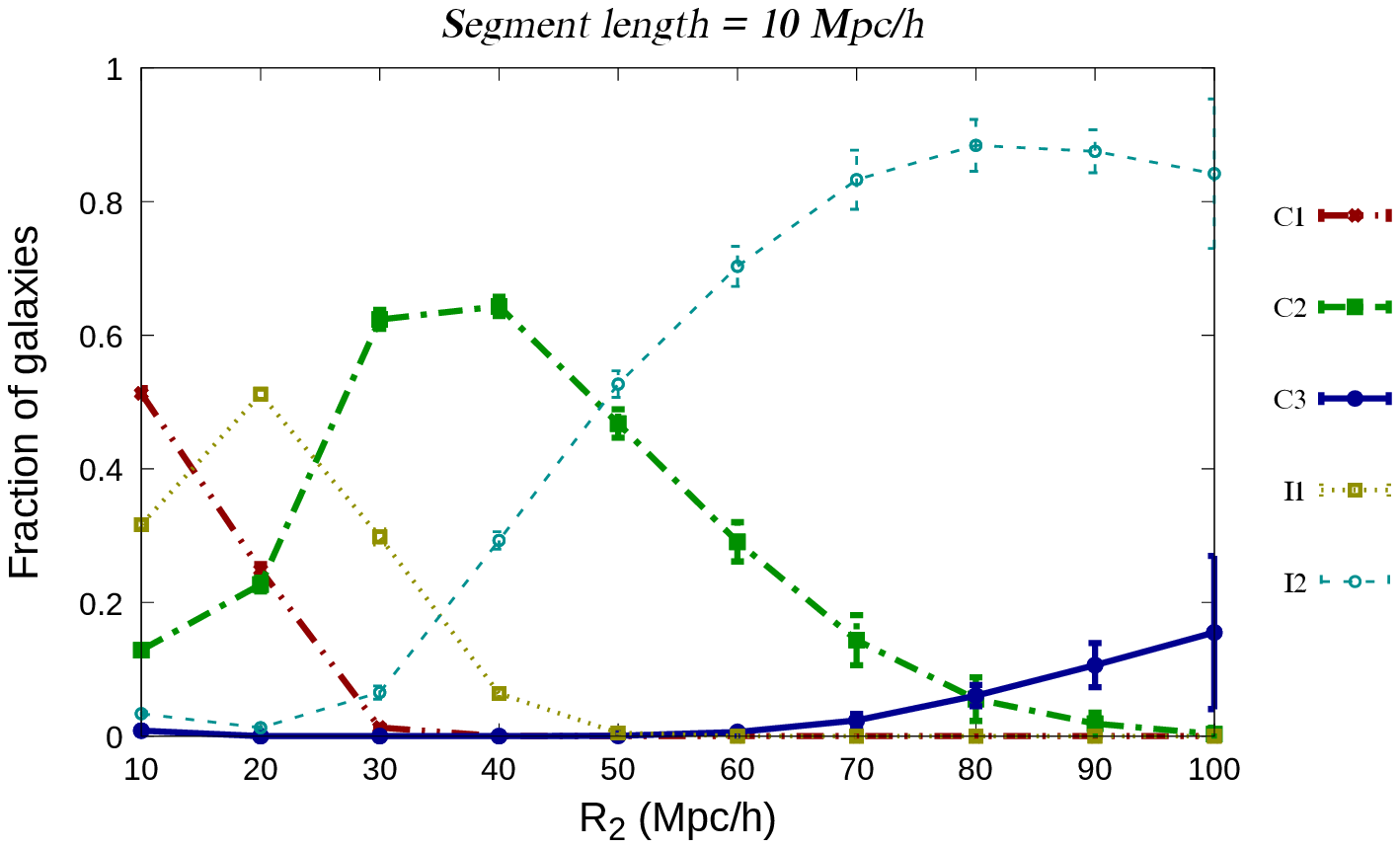}}}
\resizebox{7.5 cm}{!}{\rotatebox{0}{\includegraphics{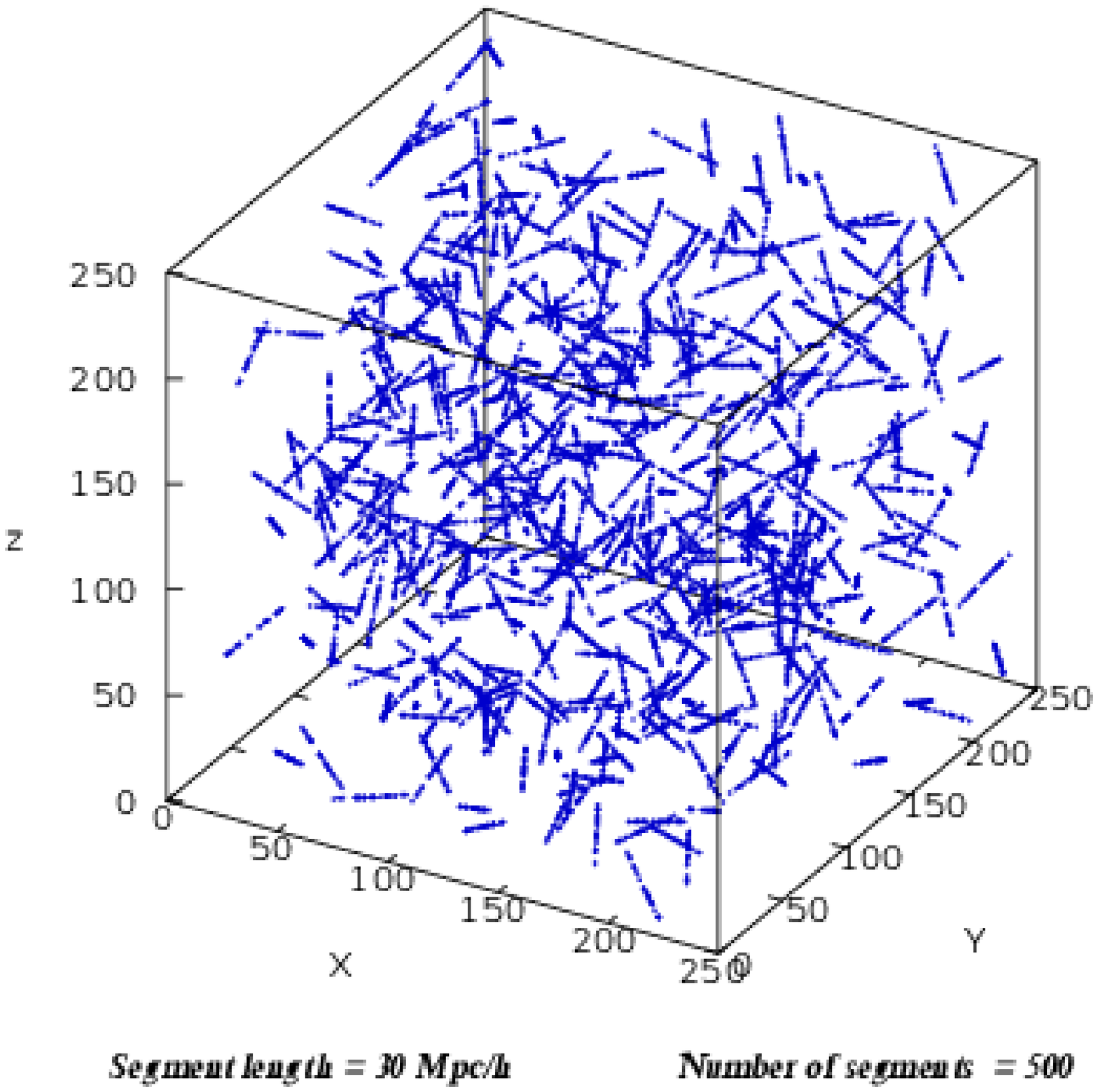}}} %
\resizebox{8.5 cm}{!}{\rotatebox{0}{\includegraphics{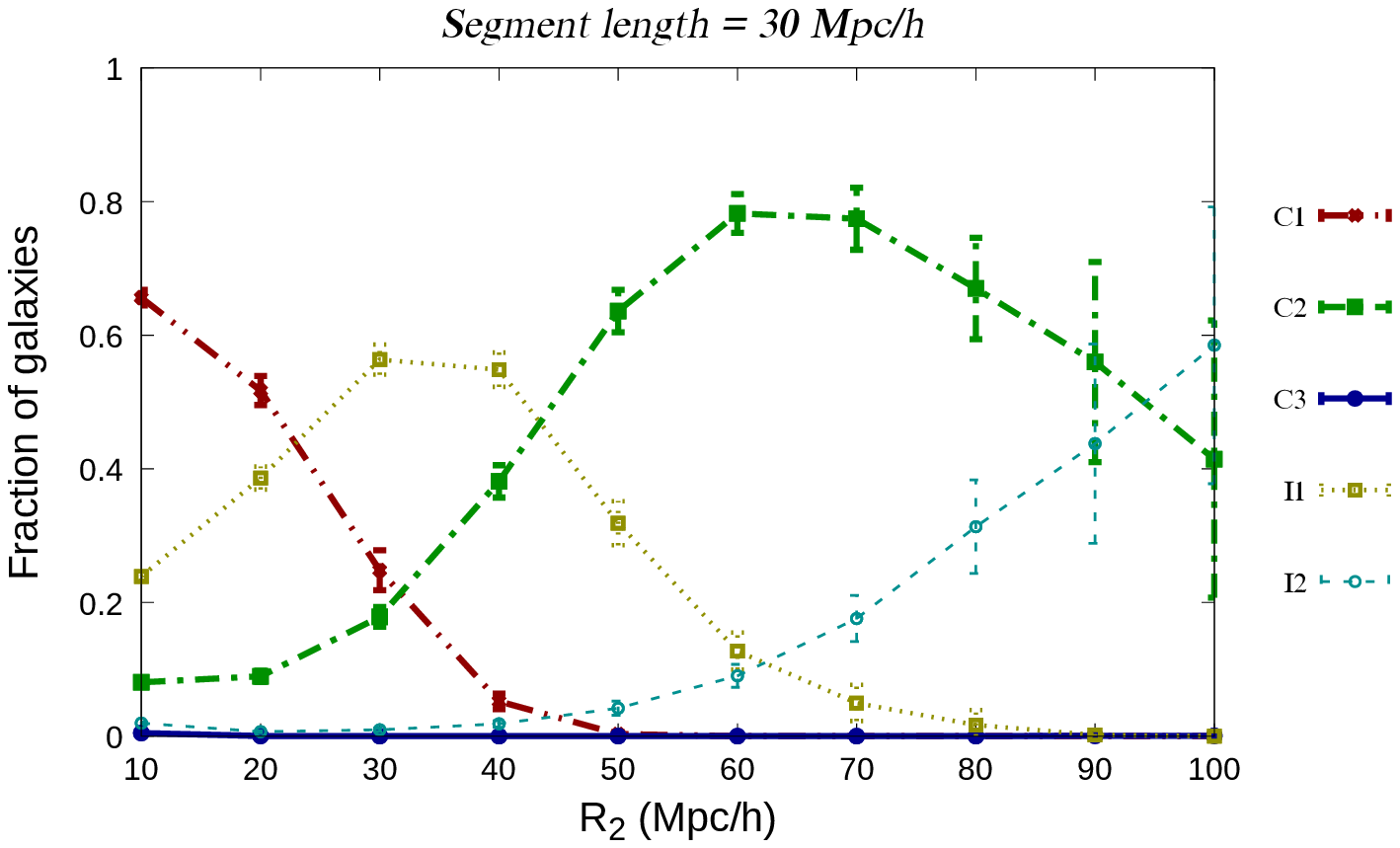}}}
\resizebox{7.5 cm}{!}{\rotatebox{0}{\includegraphics{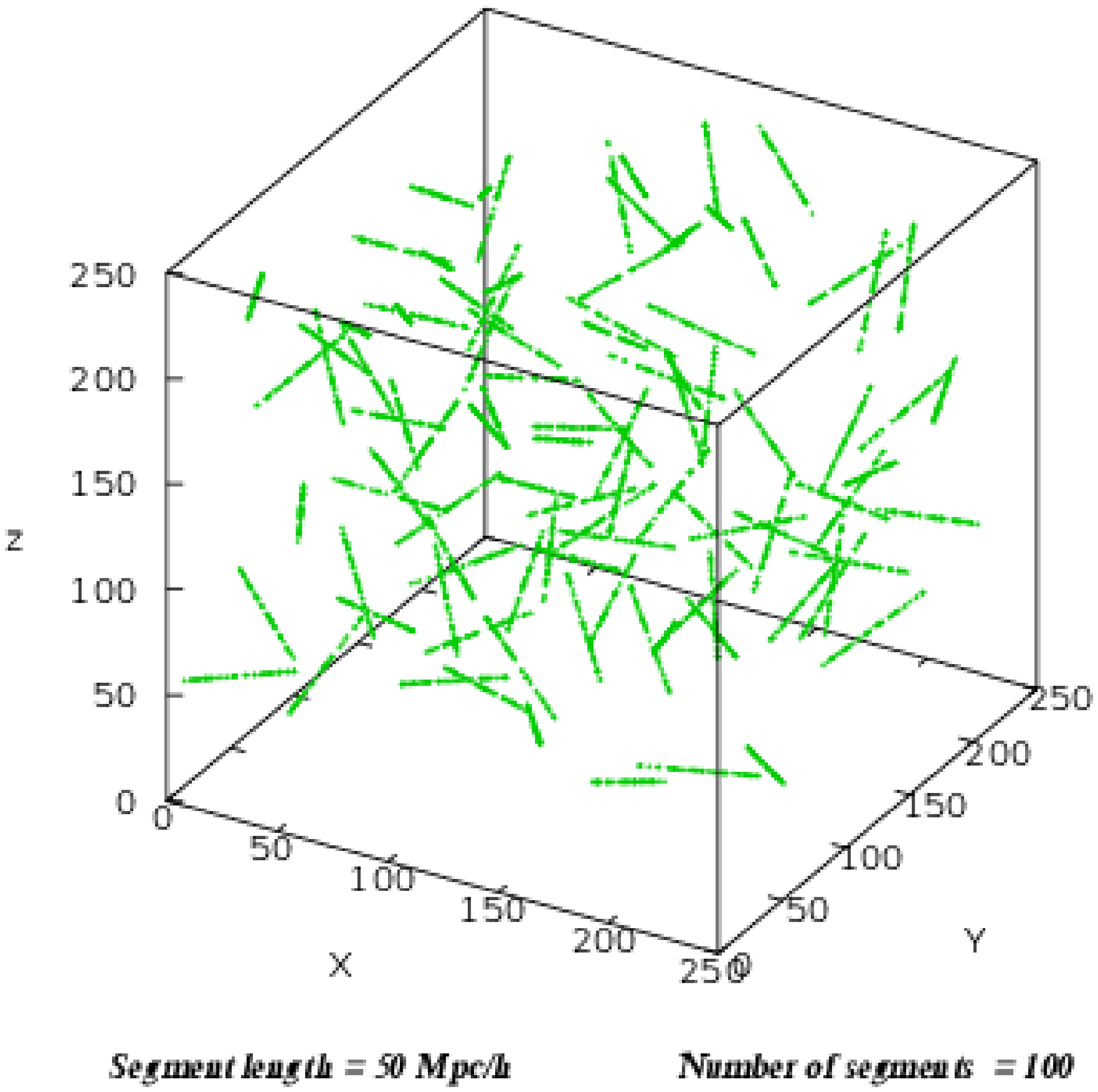}}} %
\resizebox{8.5 cm}{!}{\rotatebox{0}{\includegraphics{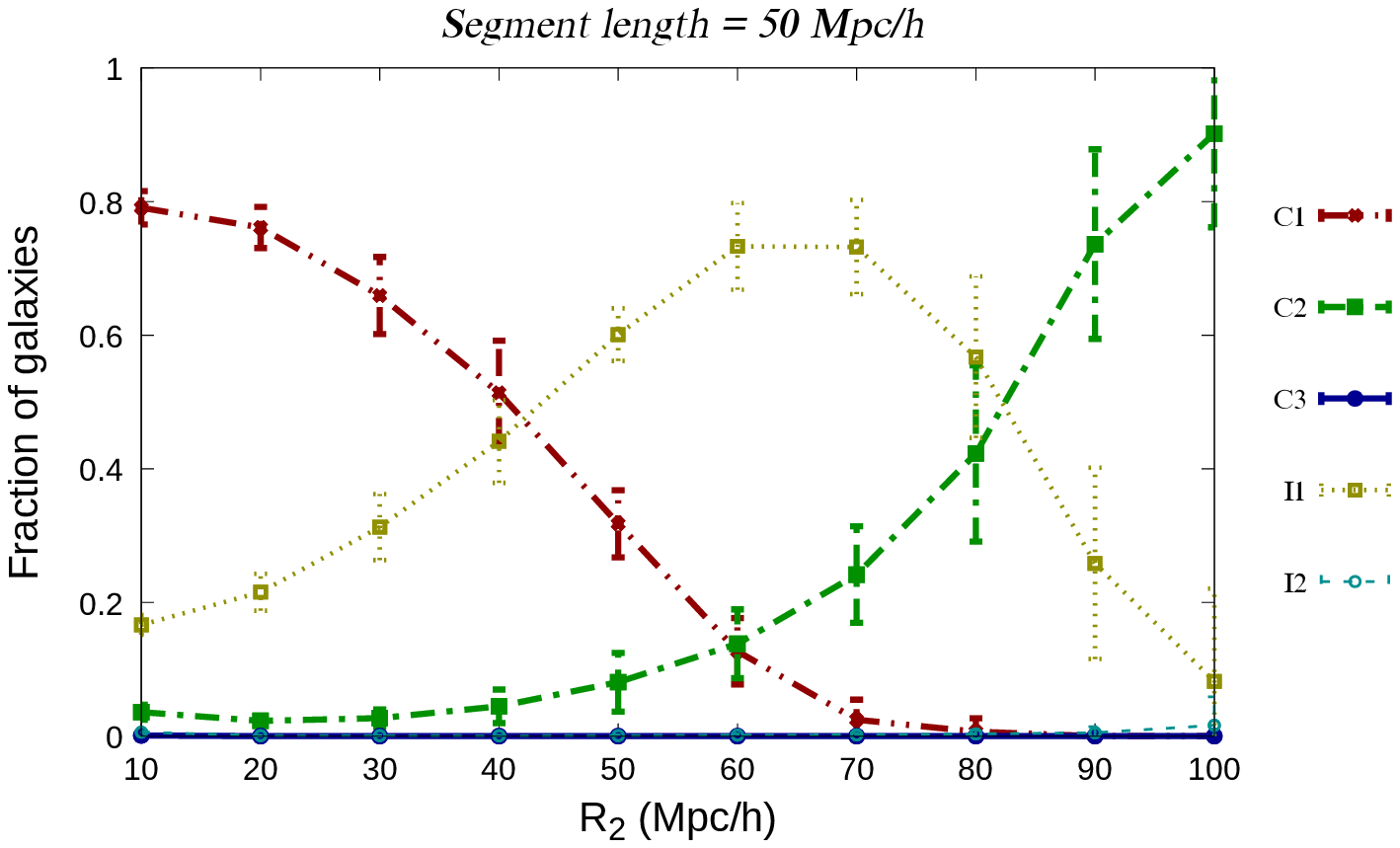}}}
\caption{The top, middle and bottom left panels of the figure show the
  3D distributions of the points generated using segment Cox process
  with segment length and number specified in each panel. The top,
  middle and bottom right panels show the fraction of points in
  different classes at different length scales for the datasets shown
  in the respective left panel.}
  \label{fig:seg_Cox}
\end{figure*}

\subsection{Scale dependence of the local dimension}

We study the scale dependence of the local dimension by identifying
the classifiable galaxies at different length scales and estimating
the local dimension for each them. We keep $R_1$ fixed at $5 \hmpc$
and gradually change $R_2$ from $10 \hmpc$ to $100 \hmpc$ in uniform
steps of $10 \hmpc$. The number of classifiable galaxies decreases
with increasing length scales. We find that initially 66136 galaxies
out of the total 90406 galaxies ($\sim 73\%$) are classifiable at
$R_2=10 \hmpc$ which decreases to 2717 ($\sim 3\%$) at $R_2=100
\hmpc$. We measure the number of galaxies classified in each class
(\autoref{tab:gclass}) and their fractions at each value of
$R_2$. The number of SDSS galaxies in each class and their fractions
as a function of $R_2$ are shown in the top left and top right panels
of \autoref{fig:nclass_fclass}. At any given length scale $R_2$, the
fractions are simply the ratio of the number of galaxies in each class
and the total number of classifiable galaxies at that length scale.

The filaments and the sheets are the most striking visible features in
the cosmic web. In our analysis, the $C1$ and $C2$ types of galaxies
are believed to be part of filament and sheet respectively. The right
panel of \autoref{fig:nclass_fclass} shows the change in the fractions
of different types of galaxies with increasing length scales. The
figure shows that at $R_2 = 10 \hmpc$, $\sim 50\%$ ($\sim 30\%$ in
sheets and $\sim 20\% $ in filaments) of all the classifiable galaxies
resides in sheets and filaments.  The rest $50\%$ galaxies are
distributed in $C3$ type and the intermediate $I1$ and $I2$ type
environment. The $C3$ type represent the galaxies inside
groups/clusters or volume filling structures such as a homogeneous
network of galaxies. The $I1$ type galaxies are expected to lie in the
vicinity where filaments and sheets intersect. Further, this may also
include the galaxies which are part of a curved or warped
filament. The $I2$ type galaxies are expected to be a part of the
environment where multiple sheets intersect. It is interesting to note
that the fraction of $C1$ type galaxies decreases rapidly from $20\%$
at $10 \hmpc$ to merely $0.1\%$ at $20 \hmpc$. We find very few $C1$
type galaxies beyond this length scale. It may be noted that only the
galaxies residing in the straight filaments would be identified as
$C1$ type. This indicates that the straight filaments do not extend
beyond a length scale of $30 \hmpc$. 

On the other hand, the fraction of $C2$ type galaxies initially
increases with length scales and peaks at $30 \hmpc$. We find that
$40\%$ classifiable galaxies are $C2$ type at $30 \hmpc$. The fraction
of $C2$ type galaxies then decreases steadily with increasing length
scales. We note that only $0.5\%$ classifiable galaxies are $C2$ type
at $90 \hmpc$. The presence of a peak at $30 \hmpc$ for the $C2$ type
galaxies indicates that most of the sheets extend upto a length scale
$30 \hmpc$. Sheets of larger sizes also exist in the cosmic web but
they become less and less abundant with increasing length scales.

The fraction of $I1$ type galaxies behaves similar to the $C1$ type
galaxies but they extend to a larger length scale. The fraction of
$I1$ type galaxies changes from $\sim 35\%$ at $10 \hmpc$ to $0.1\%$
at $60 \hmpc$ indicating that the size of such environment extend much
beyond the size of the straight filaments.  

The fraction of $I2$ type galaxies in the SDSS increases from $\sim
15\%$ at $10 \hmpc$ to $\sim 60 \%$ at $100 \hmpc$. Similarly, the
fraction of $C3$ type galaxies grows from $10\%$ at $10 \hmpc$ to $40
\%$ at $100 \hmpc$. These indicates that more and more galaxies are
associated with such environment as the length scales are increased
and nearly all the classifiable galaxies are part of either $I2$ or
$C3$ type environment on a length scale of $100 \hmpc$. This trend
clearly indicates that a nearly homogeneous network of galaxies emerge
on larger length scales.

The two middle panels of \autoref{fig:nclass_fclass} show the numbers
and fractions of different types of galaxies as a function of length
scales for the galaxies from the semi analytic galaxy catalogue from
the Millennium simulation. Interestingly, the galaxies in this semi
analytic model recovers the observed fractions of different types of
galaxies in the SDSS remarkably well. The filaments and sheets extends
upto nearly the same length scale in both the SDSS and the semi
analytic model. Some small differences in the results can be also
noted. For instance at $10 \hmpc$, a relatively higher fraction of
SDSS galaxies reside in sheets as compared those from the semi
analytic model and this trend continues till $50 \hmpc$. Further the
fraction of $C2$ type galaxies peaks at $20 \hmpc$ in the semi
analytic model whereas the same peak appears at $30 \hmpc$ for the
SDSS galaxies. This implies that the sheets are relatively less
abundant in the semi analytic model as compared to the
SDSS. Interestingly, the sheets extend upto nearly $80-90 \hmpc$ in
both the distributions. The $I2$ type galaxies which dominates the
larger length scales are also believed to inhabit the regions which are
partly sheetlike. These result emphasizes the prevalence of sheets in
the cosmic web.

It should be also noted that some of the filaments and sheets
identified in the galaxy distribution may be a result of random chance
alignment. We would like to examine this by analyzing a set of mock
SDSS catalogues from the Poisson random distributions. The results for
the Poisson distributions are shown in the bottom two panels of
\autoref{fig:nclass_fclass}. It is interesting to note that a very
small number of galaxies ($\sim 5\%$) are found inside filament at $10
\hmpc$ in the Poisson distributions. This number is roughly
$\frac{1}{4}^{th}$ of that observed in the SDSS and the semi analytic
model. This indicates that although a small number of filaments arise
due to ransom chance alignments, the majority of the filaments
detected in the SDSS and the semi analytic model are genuine in
nature. On the other hand, a significant number of galaxies ($\sim
30\%$ are of $C2$ type) in the Poisson distribution are found to be
part of a sheetlike structures at $10 \hmpc$. The fraction of both the
$C2$ type and $I1$ type galaxies diminish rapidly with increasing
length scales and becomes nearly extinct beyond $30 \hmpc$ in a
Poisson distribution. Contrary to this, we observed that the sheetlike
structures extend upto $90 \hmpc$ in the SDSS and the semi analytic
model. This suggests that the sheets identified on smaller length
scales may be a result of random chance alignment but the sheetlike
structures spanning out to larger length scales in the SDSS and the
semi analytic model are significant and genuine.

The fraction of $I2$ and $C3$ galaxies rises with increasing length
scales in both the SDSS and the semi analytic model. We note that in
the Poisson distribution, the fraction of $I2$ galaxies though
initially increases with length scales upto $30 \hmpc$ but then
decreases gradually with increasing length scales. This clearly
indicates that both sheetlike ($C2$ type) and partly sheetlike ($I2$
type) structures are less likely to emerge on larger length scales in
a Poisson distribution due to random chance alignment. This emphasizes
the significance of the large sheetlike structures observed in both
the SDSS and the semi analytic model. Finally, the fraction of $C1$
type galaxies steadily increases from $20\%$ at $10 \hmpc$ to $\sim
90\%$ at $100 \hmpc$ in the Poisson distribution indicating its
homogeneous nature as compared to the galaxy distributions on most
length scales.

\subsection{Transition of the local dimension}

We find that the galaxies tend to inhabit regions with higher local
dimension when probed on larger length scales. But most of the
galaxies which are classified according to their local dimension on
different length scales are not available at all scales. The gradual
transition of the environment of a galaxy with increasing length
scales can be only probed if its local dimension can be calculated at
each and every length scales. We identify a subset of the classifiable
of galaxies for which the local dimension can be computed throughout
the entire length scales probed. We find that there are altogether
$2282$ galaxies in our SDSS sample for which this can be achieved. We
prepare such a sample of galaxies for both the mocks from semi
analytic galaxy catalogue and Poisson distribution.

We study the variation in the fraction of different types of galaxies
as a function of length scale in each of these samples. The results
are shown in \autoref{fig:Dtrans_all}. The top left and right panel of
\autoref{fig:Dtrans_all} show the results for the SDSS and the semi
analytic model respectively. The results show that the galaxies reside
in all sorts of environment when we probe only their immediate
neighbourhood. As we include larger and larger scales in the
computation of local dimension, it appears that there are no filaments
beyond $30 \hmpc$. Interestingly, the sheetlike structures are found
to exist on length scales upto $70 \hmpc$ in both the SDSS and the
semi analytic model. The local dimension of a galaxy on length scales
beyond $70 \hmpc$ are either $I2$ type or $C3$ type indicating a
transition towards a homogeneous network. We show the results for the
Poisson distribution in the bottom middle panel of
\autoref{fig:Dtrans_all}. The results for the Poisson distribution
indicate that the filaments do not extend beyond $10 \hmpc$ and sheets
do not extend beyond $20 \hmpc$. Also there are very small number of
galaxies residing in filaments in the Poisson distribution. These
filaments and sheets are the result of random chance alignment which
should be kept in mind while analyzing any galaxy distribution to
identify various patterns present in them. These results show that
sheets can not arise from chance alignment on large scales and the
prevalence of sheetlike structures in the SDSS galaxy distribution is
an important characteristics of the observed cosmic web.

\subsection{Systematic effects}

We also study the systematics effects which may affect the outcome of
the present analysis. While estimating the local dimension, the good
quality fits are identified by employing a cut-off in the the
Chi-square per degree of freedom $\frac{{\chi}^2}{\nu} \le 0.5$. We
would like to test if the results of the present analysis are
sensitive to this criteria. We repeated our analysis for another two
cut-off values $\frac{\chi^2}{\nu} \leq 1$ and $\frac{\chi^2}{\nu}
\leq 2$. The results of this test on the SDSS data are shown in the
top two panels of \autoref{fig:systematic}. Comparing these with the
top right panel of \autoref{fig:nclass_fclass}, we find that the
fraction of galaxies available in different environment as a function
of lengthscale is insensitive to the choice of the cut-off in
$\frac{{\chi}^2}{\nu}$. We have checked that the galaxies belonging to
a particular class remains in the same class when we change the
cut-off in the $\frac{{\chi}^2}{\nu}$. It is only the numbers in each
class which gets reduced when more stringent cuts are applied.

Further, we also test if the specific number density in our volume
limited sample plays any role in deciding the results of the present
analysis. We separately repeated our analysis by randomly discarding
$25\%$ and $50\%$ of the galaxies from the SDSS volume limited sample
and adopting $\frac{{\chi}^2}{\nu} \le 0.5$. The results of this test
are shown in the middle two panels of \autoref{fig:systematic}. We
observe some small differences with the original result when $25\%$
galaxies and $50\%$ galaxies are discarded.

These tests show that the results of the present analysis are robust
and nearly independent of the cut-off in the $\frac{{\chi}^2}{\nu}$
and the number density of the galaxy sample.

The galaxy distribution analyzed here is restricted within $z<0.1385$
which probes the local Universe. In this case one may convert
redshifts to distances by simply using cosmography without the use of
any particular cosmological model. We compare the results from the
SDSS using the $\Lambda$CDM model and the cosmography in a model
independent way in the two bottom panels of \autoref{fig:systematic}.
We observe that the main findings of the analysis remain nearly model
independent.

\subsection{Tests with the segment Cox process}

We also test the efficiency of the method by simulating a set of
segment Cox process and analyzing them with the local dimension. While
analyzing the datasets from the segment Cox process, we find that the
fraction of points belonging to C1 type or the filaments type
gradually decreases with increasing length scales and extends upto a
length scale which is somewhat larger than the characteristic segment
length in each case. For example, the top right panel of
\autoref{fig:seg_Cox} shows that the fraction of filament type points
are largest ($>50\%$) at $10 \hmpc$ which decays to nearly zero at $30
\hmpc$. The datasets with segment length $30 \hmpc$ shows that more
than $65\%$ of the points are filament type at length scales of $10
\hmpc$ which gradually decays to zero at length scales of $\sim 50
\hmpc$. Similarly, we find that the dataset with segment length $50
\hmpc$ exhibit that $\sim 80\%$ points reside in filaments at $10
\hmpc$ which diminish to zero at $80 \hmpc$. The results suggest the
existence of larger number of segments having length smaller compared
to the characteristic segment length and smaller number of segments
with a length larger than the characteristic segment length. This may
arise due to the intersection and chance alignments of multiple
segments which could produce both shorter and longer segments as
compared to the characteristics segment length. The intersection of
multiple segments is more likely to occur as compared to the chance
alignment of multiple segments. This explains why we find a larger
fraction of shorter filaments than the longer filaments as compared to
the characteristic segment length. We also note that the chance
alignments of many linear segments from various orientations on large
scales can give rise to structures with sheetlike appearance. In all
the right panels of \autoref{fig:seg_Cox}, we find that a large
fraction of points are classified as C2 type or sheet type on
increasingly larger scales.  These sheetlike structure are the results
of pure chance alignments of many linear segments oriented along
different directions. Interestingly, we find that the fraction of
points belonging to volume filling structures are negligible in each
case. So the test suggests that the local dimension is unable to trace
the exact size of the linear segments in a segment Cox process but
gives the size of longest straight filaments in the distribution which
can arise after intersection and alignments of multiple segments are
taken into account. It may be noted that this increases with the
characteristic segment length. Some spurious sheetlike features are
identified on large scales due to the intersection and chance
alignments of the linear segments in the segment Cox process. However,
the fraction of C2 type points or sheetlike points remain very small
on small scales which may be used to distinguish the segment Cox
process from other types of distribution. The cosmic web is a much
more complex system than a simple superposition of linear segments of
uniform length and hence all the findings of this test may not be
applicable to the real galaxy distributions. However, the test
ascertains that the local dimension method can characterize a
distribution which is dominated by linear filamentary structures.

\section{CONCLUSIONS}

We compute the local dimension of galaxies in a volume limited galaxy
sample from the SDSS in the local Universe and study their proportions
on different length scales. We find that the galaxies reside in all
types of environments when the environment is characterized on smaller
length scales. Our results indicate that the filaments in the galaxy
distribution extend upto $30 \hmpc$ whereas the sheets extend upto as
large as $90 \hmpc$. On large scales, the majority of the galaxies in
the SDSS are found to reside in either sheetlike or partly sheetlike
environment. We find a very similar trend in the semi analytic galaxy
catalogue from the Millennium Run simulation. No filaments and sheets
are observed beyond a length scales of $30 \hmpc$ in the Poisson
distribution. The absence of sheetlike structures on large scales in
the Poisson distribution show that they can not result from a random
chance alignment on those length scales. The present analysis find the
prevalence of sheetlike structures in the cosmic web on larger length
scales. The filaments are only observed on smaller length scales and
are completely absent on larger length scales.

Our analysis indicates that the sheets and the sheetlike structures
are the most dominant features on large scales in the galaxy
distribution from the SDSS as well as in the semi analytic model. In
the Zeldovich scenario, the pancakes are the first non-linear
structures formed by gravitational collapse. \citet{dorosh} show that
the simultaneous collapse along multiple axes is quite unlikely and
the filaments and nodes would form later depending on the eigenvalues
of the deformation tensor at different Lagrangian co-ordinates. The
pancakes are expected to be the most dominant feature emerging from
the first stage of non-linear clustering. So the higher abundance of
sheets or sheetlike structures observed on relatively larger scales
may be a consequence of the Zeldovich approximation.

Earlier studies find that the filaments are statistically significant
upto length scales of $70 \hmpc$ \citep{pandey1} whereas our results
indicate that the straight filaments can only extend upto $30 \hmpc$.
The two dimensional sections analyzed by \citet{pandey1} may also
include some filaments which arise due to the projection of sheetlike
structures. Further, their study also incorporates the curved or
wiggly filaments into consideration.

The observed galaxy distribution shows a tendency towards transition
to a homogeneous network on larger length scales. This is consistent
with the findings that the Universe is homogeneous around a length
scales of $\sim 100 \hmpc$ \citep{yadav, hogg, prakash, scrim,
  nadathur, pandey15, pandey16, avila}.

We study the systematics effects of the number density of the sample
and the cut-off in the goodness of fit and find that our results are
robust against the variation in these parameters. Analyzing simulated
datasets of the segment Cox process, we find that the local dimension
method can characterize such distributions.

\section{ACKNOWLEDGEMENT}
The authors thank an anonymous reviewer for useful comments and
suggestions which helped us to improve the draft. The authors would
like to thank the SDSS team for making the data public. SS would like
to thank UGC, Government of India for providing financial support
through a Rajiv Gandhi National Fellowship.  B.P. would like to
acknowledge financial support from the SERB, DST, Government of India
through the project EMR/2015/001037. B.P. would also like to
acknowledge IUCAA, Pune and CTS, IIT, Kharagpur for providing support
through associateship and visitors programme respectively.

\bsp	
\label{lastpage}
\end{document}